\documentclass[12pt,a4paper]{article}
\usepackage{a4}
\usepackage[dvips]{epsfig}
\usepackage{amssymb}
\newlength{\defbaselineskip}
\newcommand{\setlinespacing}[1]%
            {\setlength{\baselineskip}{#1 \defbaselineskip}}

\begin{document}

\title{\textbf{\emph{Ab initio} calculations of the atomic and electronic
structures for ABO$_{3}$ perovskite (001) surfaces}}

\author{S. Piskunov$^{a,b}$, E. A. Kotomin$^{b}$, E.
Heifets$^{c}$}

\date{\small{May 7, 2004}}

\maketitle

\vspace*{-0.75cm} \noindent \centerline {\small $^{a}$Fachbereich
Physik, Universit\"{a}t Osnabr\"{u}ck, D-49069 Osnabr\"{u}ck,
Germany} \noindent \centerline {\small $^{b}$Institute of Solid
State Physics, University of Latvia, Kengaraga 8, LV- 1063 Riga,
Latvia} \noindent \centerline {\small $^{c}$California Institute
of Technology, MS 139-74, Pasadena CA 91125, USA} \vspace*{0.5cm}

\begin{abstract}
\noindent We present the results of first-principles calculations
on two possible terminations of the (001) surfaces of SrTiO$_{3}$
(STO), BaTiO$_{3}$ (BTO), and PbTiO$_{3}$ (PTO) perovskite
crystals. Atomic structure and the electronic configurations
were calculated for different 2D slabs, both stoichiometric and
non-stoichiometric, using \emph{hybrid} (B3PW)
exchange-correlation technique and reoptimized basis sets of
atomic (Gaussian) orbitals. Results are compared with previous
calculations and available experimental data. The electronic
density distribution near the surface and covalency effects are
discussed in details for all three perovskites.
\end{abstract}


\def\baselinestretch{1.5}
\setlinespacing{1.11}

\section{Introduction}

The ABO$_3$ perovskite surfaces are important for many
technological applications, \emph{e.g.} STO is widely used as a
substrate for epitaxial growth of high-\emph{T$_{c}$}
superconducting thin films, BTO and PTO and their solid solutions
are promising for non-volatile memory cells, as electro-optical
materials, and for piezoelectrical devices
\cite{linesb,noguerab,henricb,scott-feram}. Due to intensive
development and progressive miniaturization of electronic devices,
the electronic properties and atomic structure of the ABO$_{3}$
perovskite thin films were extensively studied during the
last years. The STO(001) surface structure was analyzed by
means of Low Energy Electron Diffraction (LEED)
\cite{bickel-PRL62}, Reflection High Energy Electron Diffraction
(RHEED), X-ray Photoelectron Spectroscopy (XPS) and Ultraviolet
Electron Spectroscopy (UPS) \cite{hikita-ss287}, Medium Energy Ion
Scattering (MEIS) \cite{ikeda-SS433}, and Surface X-ray
Diffraction (SXRD) \cite{charlton-SS457}. The most recent
experimental studies on the STO(001) include a combination of XPS,
LEED, and Time-Of-Flight Scattering and Recoiling Spectrometry
(TOF-SARS) \cite{heide-SS473}, and UPS and Metastable Impact
Electron Spectroscopy (MIES) \cite{kempter-SS515}. The BTO and PTO
surfaces are less studied. The first \emph{ab initio} calculations
of the ABO$_3$ perovskite surfaces were presented by Kimura
\emph{et al.} in 1995 \cite{kimura-prb51}. Since then a number of
calculations were performed, using different methods, \emph{e.g.}
Linearized Augmented Plane Waves (LAPW)
\cite{cohen-jpcs57,cohen-frr194}, PW ultrasoft-pseudopotential
\cite{vand-prb56,vand-ss418,Vand-far114,Xue-ss550}, Hartree-Fock
(HF) \cite{cora-catFD}. Recently, new studies on the STO(001)
surface relaxation were performed using Density Functional
Theory (DFT) with plane wave (PW) basis set (DFT-PW method)
\cite{Cheng-prb62} and Shell Model (SM) simulations with
parameters obtained from first-principles LAPW method
\cite{Tinte-aip535,Tinte-prb64}. In most of these studies DFT and
plane-wave basis sets were used (the only exception is HF study
\cite{cora-catFD}). It is well-known that DFT considerably
underestimate the band gap. On the other hand, band gap obtained
through the HF calculations is usually an overestimate
\cite{Pisani-ed-96}. To solve this problem, the functionals
containing \emph{``hybrid"} of the non-local HF exchange, DFT
exchange, and Generalized Gradient Approximation (GGA) correlation
functionals, and known as B3LYP and B3PW (which are quite popular
in quantum chemistry of molecules)
\cite{curtiss-JCP106,curtiss-JCP109} could be used. Recently,
periodic-structure \emph{ab initio} hybrid calculations were
carried out for wide range of crystalline materials
\cite{harrison-b3lyp}, as well as for perovskites and their
surfaces \cite{pisk-BS,heif-ss02}. In all cases the hybrid
functional technique shows the best agreement with experimental
data for both bulk geometry and optical properties of materials
under investigation. In this paper we continue our own theoretical
investigation of the surfaces of perovskite materials. These have
been carried out using both semi-empirical SM \cite{heifss2000}
and \emph{ab initio} HF and DFT methods
\cite{heif-prb01,kotomin-TSF400,heif-ss02,borstel-PSSb236}.
Mostly, these studies are dedicated to STO(001) surfaces. We
studied the effects of different type of Hamiltonians (varied from
DFT-LDA to HF with \emph{posteriori} corrections) on surfaces
properties and atomic structure. This analysis allowed us to
choose the B3PW functional as the best for describing the  bulk
properties of all three perovskites under consideration
\cite{pisk-BS} as well as the basic surface properties
\cite{heif-ss02}. This is why in the present study we use B3PW for
comparing the atomic structure and the electronic properties of
the (001) surfaces of three similar perovskites, STO, BTO, and
PTO. In our simulations, the (001) surfaces of perovskites are
calculated using a \emph{slab} model.

The paper is organized as follows: in Section \textbf{2} we
present the technical details of calculations and methodology. In
Section \textbf{3} we discuss the surface structure and the
electronic properties of the (001) surfaces. Our conclusions are
summarized in Section \textbf{4}.

\section{Computational Details}

To perform the first-principles DFT-B3PW calculations, the
CRYSTAL'98 computer code
\cite{Pisani-ed-96,CRman,CR-http1,CR-http2} was used. This code
uses localized Gaussian-type functions (GTF) localized at atoms as
the basis for an expansion of the crystalline orbitals. The
ability to calculate the electronic structure of materials within
both HF and Kohn-Sham (KS) Hamiltonians and implementation of
purely 2D slab model without its artificial repetition along the
\emph{Z} axis, are the main advantages of this code. However, in
order to employ the LCAO-GTF method, it is desirable to optimize
the basis sets (BS), which would be suitable for the electronic
structure computations. Such BS's optimization for all three
perovskites is developed and discussed in Ref. \cite{pisk-BS}.
Unlike the standard basis set \cite{CR-http1}, we added the
polarized O $d$ orbitals, replaced the Ti inner core orbitals by
the small-core Hay-Wadt presudopotentials, and consistently used
the two diffuse $s$ and $p$ Gaussians as the separate basis
orbitals on the Ti, Ba, Sr and Pb.

Our calculations were performed using the hybrid
exchange-correlation B3PW functional involving a hybrid of
non-local Fock exact exchange and Becke's three-parameter gradient
corrected exchange functional \cite{becke-hybr} combined with the
non-local gradient corrected correlation potential by Perdew and
Wang \cite{PW1,PW2,PW3}. The Hay-Wadt small-core ECP's
\cite{hw1,hw2,hw3} were adopted for Ti, Sr, and Ba atoms
\cite{hw1,hw2,hw3}. The ``small-core" ECP's replace only inner
core orbitals, but orbitals for outer core electrons as well as
for valence electrons are calculated self-consistently. Light
oxygen atoms were left with the full electron BS. The BSs were
adopted in the following forms: O - 8-411(1d)G (the first shell is
of \emph{s}-type and is a contraction of eight Gaussian type
functions, then there are three \emph{sp}-shells and one
\emph{d}-shell), Ti - 411(311d)G, Sr and Ba - 311(1d)G; see Ref.
\cite{pisk-BS} for more details.

The reciprocal space integration was performed by the sampling the
Brillouin zone of the unit cell with the $8\times8\times1$
Pack-Monkhorst net \cite{monkhorst}, that provides the balanced
summation in direct and reciprocal lattices
\cite{bredov-evarestov}. To achieve high accuracy, large enough
tolerances of 7, 8, 7, 7, 14, (i.e. the calculation of integrals
with an accuracy of 10$^{-N}$) were chosen for the Coulomb
overlap, Coulomb penetration, exchange overlap, the first exchange
pseudo-overlap, and for the second exchange pseudo-overlap
respectively \cite{CRman}.

\setlength{\fboxsep}{1pt}
\begin{figure}[htbp]
\def\baselinestretch{1.0}
\setlinespacing{1.0}
  \begin{center}
      \epsfig{file=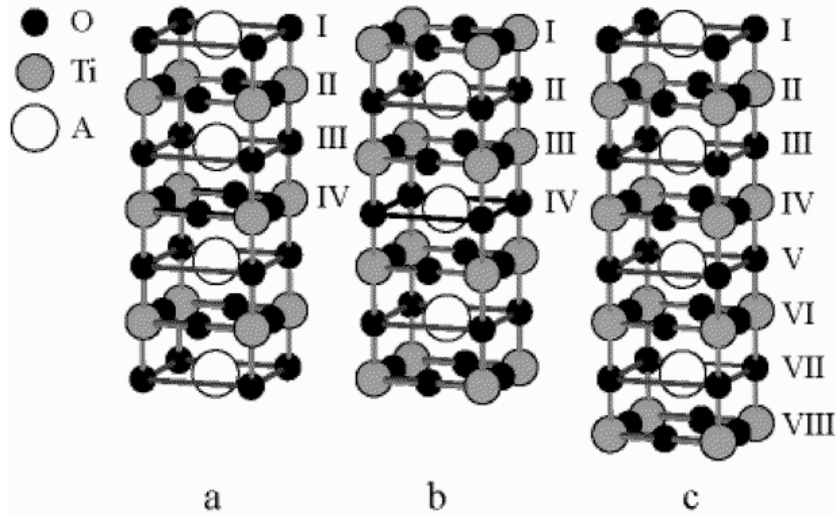,angle=0,width=11cm}
    \caption{
\small Schematic view of the modelling for ABO$_3$(001) surfaces:
a) AO-terminated on both sides, b) TiO$_{2}$-terminated on both
sides, c) asymmetrical termination (AO atop and TiO$_2$ on
bottom).
      }
    \label{surfaces}
  \end{center}
\end{figure}
The ABO$_3$(001) surfaces were modelled considering the crystal as
a set of crystalline planes perpendicular to the given surface,
and cutting out 2D slab of a finite thickness, periodic in
\emph{x-y} plane. The symmetrical 7-plane slabs are either AO- or
TiO$_2$-terminated. These are non-stoichiometric with unit cell
formulas A$_4$B$_3$O$_{10}$ and A$_3$B$_4$O$_{11}$, respectively.
Asymmetrical slab is AO- and TiO$_2$-terminated (from each side,
respectively), and stoichiometric with a formula
A$_4$B$_4$O$_{12}$ unit cell -- are shown schematically in Fig.
\ref{surfaces}. These slabs containing seven planes (symmetric)
and eight planes (asymmetric) can be considered as thick enough
since the convergence of calculated slab total energy per ABO$_3$
unit is achieved. This energy differs less than $5\cdot 10^{-4}$
Hartree for 7- and 9-layered (or 8- and 10-layered for
asymmetrical termination) slabs for all three perovskites. The
symmetric slabs have advantage of no dipole moment perpendicular
to the slab. However, their non-stoichiometry can potentially
affect the electronic density distributions and thus atomic
displacements. On the other hand, since the effective atomic
charges of Ti and O ions differ from the ionic charges +4e, -2e
(due to the covalency contribution in Ti-O chemical bonding), the
alternating TiO$_2$ and AO planes are slightly charged, which
produces in the asymmetrical slab certain dipole moment
perpendicular to the slab. This dipole moment is cancelled by the
electronic density redistribution near the surface which also can
affect the optimized atomic displacements. This is why a critical
comparison of these two slab models is important for making
reliable conclusions.

In order to compare surface properties of three perovskites, we
study here their high-symmetry cubic (\emph{Pm3m}) phases. The
calculated bulk lattice constants (in {\AA}) are: $a=3.90$ for
STO, $a=4.01$ for BTO, and $a=3.93$ for PTO, which demonstrates
the excellent  agreement with experiment \cite{springIII,shirane}.

\section{Results and Discussions}
\subsection{Surface Atomic Structure}

In present surface structure simulations we allowed atoms of two
outermost surface layers to relax along the $z$ axis (by symmetry,
surfaces of perfect cubic crystals have no forces along the $x$-
and $y$-axes). Displacements of third layer atoms were found
negligibly small in our calculations and thus are not treated. The
optimization of atomic coordinates was done through the slab total
energy minimization using our own computer code which implements
Conjugated Gradients optimization technique \cite{num-rec-f77}
with a numerical computation of derivatives.

\setlength{\fboxsep}{1pt}
\begin{figure}[htbp]
\def\baselinestretch{1.0}
\setlinespacing{1.0}
  \begin{center}
    \epsfig{file=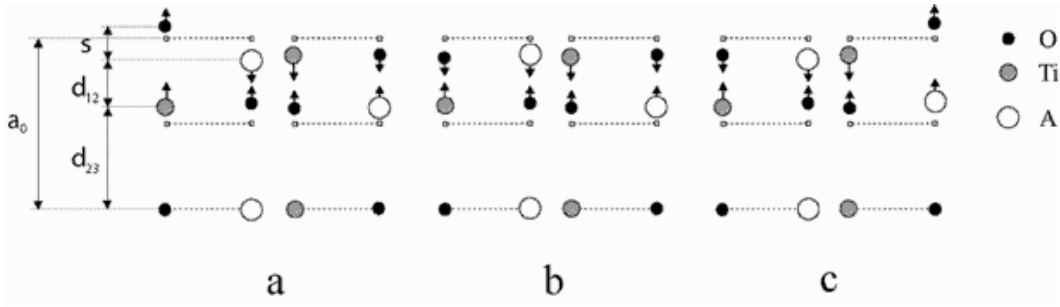,angle=0,width=14cm}
    \caption{
\small Schematic illustration of the relaxation for two outermost
surface layers relaxation: a) STO, b) BTO, c) PTO.  The view in
the [010] direction. Left panels in $a$-$c$ are AO termination,
right panels -- TiO$_{2}$. Dashed lines represent positions of the
same layers in perfect bulk crystals.
      }
    \label{relax}
  \end{center}
\end{figure}
Our calculated atomic displacements are presented in Table
\ref{surf-relax} and are schematically illustrated in Fig.
\ref{relax}. A comparison with the surface atomic displacements
obtained by other Quantum Mechanical (QM) calculations is also
done in Table \ref{surf-relax}. The relaxation of surface metal
atoms is much larger than that of oxygens which leads to a
considerable \emph{rumpling} of the outermost plane. Atoms of the
first surface layer relax inwards, \emph{i.e.} towards the bulk.
The two exceptions are the top oxygens of STO SrO-terminated and
PTO TiO$_2$-terminated surfaces (the latter PW-pseudopotential
calculations are in disagreement \cite{Vand-far114}, however the
magnitudes of both displacements are relative small, 0.31 and
-0.34 per cents of lattice constant, respectively.) The outward
relaxation of all atoms in the second layer is found for all three
perovskites and both terminations. The displacements obtained for
asymmetrically terminated slabs are practically the same as for
those symmetrically terminated. This means that both slab models
are reliable for the calculations of the (001) neutral surfaces
and that 7-8 plane slabs thick enough.
\begin{table}
\renewcommand{\baselinestretch}{1.0}
 \centering
 \caption{\small Atomic displacements with respect to atomic positions on unrelaxed ABO$_3$(001)
 surfaces (in percent of  bulk lattice constant). Symbol A can mean Sr, Ba, or Pb atom. Positive
 displacements are outwards (to the vacuum), negative displacements mean inward the slab center.}
 \label{surf-relax}
\renewcommand{\baselinestretch}{1.2}
\small\normalsize\scriptsize
\begin{tabular}{ccccccc@{\hspace{1mm}}cccccc@{\hspace{1mm}}ccc}
                           &        &          &        &        &        &        & &        &        &        &        &        & &       &        \\
  \hline
  \hline
                         N & At.    & STO      &        &        &        &        & & BTO    &        &        &        &        & & PTO   &        \\
 \cline{3-7} \cline{9-13} \cline{15-16}
                           &        & This     & \cite{heif-ss02} & \cite{heifss2000} & \cite{vand-ss418} & \cite{Cheng-prb62} & & This   & \cite{heifss2000} & \cite{vand-prb56} & \cite{Tinte-aip535} & \cite{cohen-frr194} & & This  & \cite{Vand-far114} \\
                           &        & study    &        &        &        &        & & study  &        &        &        &        & & Study &        \\
\hline
\multicolumn{16}{c}{AO-termination}\\
\hline
                         1 & A      & -4.84    & -4.29  & -7.10  & -5.7   & -6.66  & & -1.99  & -3.72  & -2.79  & -0.72  &        & & -3.82 & -4.36  \\
                           & O      & 0.84     & 0.61   & 1.15   & 0.1    & 1.02   & & -0.63  & 1.00   & -1.40  & -1.09  &        & & -0.31 & -0.46  \\
                         2 & Ti     & 1.75     & 1.25   & 1.57   & 1.2    & 1.79   & & 1.74   & 1.25   & 0.92   & 1.70   &        & & 3.07  & 2.39   \\
                           & O$_{2}$& 0.77     & 0.85   & 0.87   & 0.0    & 0.26   & & 1.40   & 0.76   & 0.48   & 2.75   &        & & 2.30  & 1.21   \\
                         3 & A      &          &        & -1.42  & -1.2   & -1.54  & &        & -0.51  & 0.53   & -0.69  &        & &       & -1.37  \\
                           & O      &          &        & 0.7    & -0.1   & 0.26   & &        & 0.16   & 0.26   & -0.28  &        & &       & -0.20  \\
\hline
\multicolumn{16}{c}{TiO$_{2}$-termination}\\
\hline
                         1 & Ti     & -2.25    & -2.19  & -2.96  & -3.4   & -1.79  & & -3.08  & -2.72  & -3.89  & -4.14  &        & & -2.81 & -3.40  \\
                           & O$_{2}$& -0.13    & -0.93  & -1.73  & -1.6   & -0.26  & & -0.35  & -0.94  & -1.63  & -2.74  &        & & 0.31  & -0.34  \\
                         2 & A      & 3.55     & 2.18   & 3.46   & 2.5    & 4.61   & & 2.51   & 2.19   & 1.31   & 2.36   &        & & 5.32  & 4.53   \\
                           & O      & 0.57     & 0.01   & -0.21  & -0.5   & 0.77   & & 0.38   & -0.17  & -0.62  & -0.50  &        & & 1.28  & 0.43   \\
                         3 & Ti     &          &        & -0.6   & -0.7   & -0.26  & &        & -0.33  & -0.75  & -0.81  &        & &       & -0.92  \\
                           & O$_{2}$&          &        & -0.29  & -0.5   & 0.26   & &        & -0.01  & -0.35  & -0.72  &        & &       & -0.27  \\
\hline
\multicolumn{16}{c}{asymmetrical termination}\\
\hline
                         1 & A      & -5.22    &        &        &        &        & & -2.09  &        &        & -1.8   & -4.28  & & -4.02 &        \\
                           & O      & 0.39     &        &        &        &        & & -0.69  &        &        & -1.98  & -3.26  & & -0.24 &        \\
                         2 & Ti     & 1.55     &        &        &        &        & & 1.70   &        &        &        &        & & 3.01  &        \\
                           & O$_{2}$& 0.61     &        &        &        &        & & 1.50   &        &        &        &        & & 1.95  &        \\
                        ...& ...    & ...      &        &        &        &        & & ...    &        &        & ...    & ...    & & ...   &        \\
                        ...& ...    & ...      &        &        &        &        & & ...    &        &        & ...    & ...    & & ...   &        \\
                        ...& ...    & ...      &        &        &        &        & & ...    &        &        & ...    & ...    & & ...   &        \\
                         7 & O      & -0.64    &        &        &        &        & & -0.37  &        &        &        &        & & -1.22 &        \\
                           & A      & -3.74    &        &        &        &        & & -2.54  &        &        &        &        & & -5.44 &        \\
                         8 & O$_{2}$& 0.15     &        &        &        &        & & 0.34   &        &        & 3.52   & 2.68   & & -0.74 &        \\
                           & Ti     & 2.27     &        &        &        &        & & 3.03   &        &        & 4.72   & 4.79   & &  2.84 &        \\
  \hline
  \hline
\end{tabular}
\end{table}

In order to compare the calculated surface structures with
available results obtained experimentally, the surface rumpling
$s$ (the relative displacement of oxygen with respect to the metal
atom in the surface layer) and the changes in interlayer distances
$\Delta d_{12}$ and $\Delta d_{23}$ ($1$, $2$, and $3$ are the
numbers of near-surface layers) are presented in Table
\ref{rumpling}. Our calculations of the interlayer distances are
based on the positions of relaxed \emph{metal} ions (Fig.
\ref{relax}), which are known to be much stronger electron
scatters than oxygen ions \cite{bickel-PRL62}. The agreement is
quite good for all theoretical methods, which give the same sign
for both the rumpling and change of interlayer distances. The
amplitude of surface rumpling of SrO-terminated STO is predicted
much larger than that for TiO$_2$-terminated STO surface, whereas
the rumpling of BTO TiO$_2$-terminated surface is predicted to
exceed by a factor of two that for BaO-terminated surface. Lastly,
PTO demonstrates practically equal rumpling for both terminations.
\begin{table}
\renewcommand{\baselinestretch}{1.0}
 \centering
 \caption{
          \small Calculated and experimental surface rumpling $s$, and relative displacements of the
          three near-surface planes for the AO- and TiO$_{2}$-terminated
          surfaces $\Delta d_{ij}$ (in percent of lattice constant).
          Results for asymmetrical surfaces are given in brackets.
          }
 \label{rumpling}
\renewcommand{\baselinestretch}{1.2}
\small\normalsize\scriptsize
\begin{tabular}{clcccccccccc}
      &                         &            &             &            & &             &             &            \\
  \hline
  \hline
      &                         & \multicolumn{3}{c}{AO-terminated}     & & \multicolumn{3}{c}{TiO$_{2}$-terminated}\\
  \cline{3-5} \cline{7-9}
      &                         & $s$ & $\Delta d_{12}$ &$\Delta d_{23}$& & $s$ & $\Delta d_{12}$ & $\Delta d_{23}$ \\
  \hline
 STO  & This study              &5.66        &-6.58        & 1.75       & &2.12         &-5.79        &3.55        \\
      &                         & (5.61)     & (-6.79)     & (1.55)     & & (2.43)      &  (-6.02)    &   (3.74)   \\
      &\emph{ab initio} \cite{heif-ss02}& 4.9 & -5.5       &            & & 1.3         & -4.4        &            \\
      &\emph{ab initio} \cite{vand-ss418}& 5.8& -6.9       & 2.4        & & 1.8         & -5.9        & 3.2        \\
      &\emph{ab initio} \cite{Cheng-prb62}&7.7& -8.6       & 3.3        & & 1.5         & -6.4        & 4.9        \\
      &Shell model \cite{heifss2000}& 8.2    & -8.6        & 3.0        & & 1.2         & -6.4        & 4.0        \\
      &LEED expt. \cite{bickel-PRL62}&4.1$\pm$2&-5$\pm$1   & 2$\pm$1    & & 2.1$\pm$2   & 1$\pm$1     & -1$\pm$1   \\
      &RHEED expt. \cite{hikita-ss287}& 4.1  & 2.6         & 1.3        & & 2.6         & 1.8         & 1.3        \\
      &MEIS expt. \cite{ikeda-SS433}  &      &             &            & & 1.5$\pm$0.2 & 0.5$\pm$0.2 &            \\
      &SXRD expt. \cite{charlton-SS457}&1.3$\pm$12.1&-0.3$\pm$3.6&-6.7$\pm$2.8& &12.8$\pm$8.5&0.3$\pm$1&           \\
  \hline
 BTO  & This study              & 1.37       & -3.74       & 1.74       & & 2.73        & -5.59       & 2.51       \\
      &                         & (1.40)     & (-3.79)     & (1.70)     & & (2.69)      & (-5.57)     & 2.54       \\
      &\emph{ab initio} \cite{vand-prb56}&1.39& -3.71      & 0.39       & & 2.26        & -5.2        & 2.06       \\
      &Shell model \cite{Tinte-aip535}&0.37& -2.42    & 2.39       & & 1.4         & -6.5        & 3.17       \\
      &Shell model \cite{heifss2000}& 4.72   & -4.97       & 1.76       & & 1.78        & -4.91       & 2.52       \\
  \hline
 PTO  & This study              & 3.51       & -6.89       & 3.07       & & 3.12        & -8.13       & 5.32       \\
      &                         & (3.78)     & (-7.03)     & (3.01)     & & (3.58)      & (-8.28)     & (5.44)     \\
      &\emph{ab initio} \cite{Vand-far114}&3.9& -6.75      & 3.76       & & 3.06        & -7.93       & 5.45       \\
  \hline
  \hline
\end{tabular}
\end{table}
From Table \ref{rumpling} one can see that all surfaces show the
reduction of interlayer distance $d_{12}$ and expansion of
$d_{23}$. The calculated surface rumpling agrees quite well with
LEED, RHEED and MEIS experiments
\cite{bickel-PRL62,hikita-ss287,ikeda-SS433} (which are available
for STO surfaces only). Theory agrees semi-quantitatively also
with the LEED results for the $\Delta d_{12}$ and $\Delta d_{23}$.
However, from Table \ref{rumpling} is well seen that LEED and
RHEED experiments contradict each other in the sign of $\Delta
d_{12}$ for SrO-terminated surface and $\Delta d_{23}$ of
TiO$_2$-terminated surface. Another problem is that LEED, RHEED
and MEIS experiments demonstrate that the topmost oxygen always
move outwards the surfaces whereas all calculations predict for
the TiO$_2$-terminated STO surface that oxygen goes
\emph{inwards}. Even more important is the contradiction between
the three above-mentioned experiments and recent SXRD study
\cite{charlton-SS457} where oxygen atoms are predicted to move
inwards for \emph{both} surface terminations reaching rumpling up
to 12.8\% for the TiO$_2$ terminated surface! Up to now the reason
for such discrepancies between the different experimental data is
not clear (see \cite{charlton-SS457,vand-ss418}). Thus, the
disagreement between data obtained theoretically and
experimentally cannot be analyzed until the conflict between
different experimental results is resolved.

\begin{table}
\renewcommand{\baselinestretch}{1.0}
 \centering
 \caption{
          \small Calculated surface energies (in eV per surface cell).
          Results for previous \emph{ab initio} calculations
          \cite{Vand-far114,Cheng-prb62,Tinte-prb64}
          are averaged over AO- and TiO$_{2}$-terminated surfaces.
          }
\label{surf-en}
\renewcommand{\baselinestretch}{1.2}
\small\normalsize\scriptsize
\begin{tabular}{cccccccccccc}
            &       &       &       & &       &       &       & &       &       &       \\
  \hline
  \hline
            & STO   &       &       & & BTO   &       &       & & PTO   &       &       \\
 \cline{2-4} \cline{6-8} \cline{10-12}
            &SrO&TiO$_{2}$&asymm.& &BaO&TiO$_{2}$&asymm.& &PbO&TiO$_{2}$&asymm. \\
 \hline
 This study & 1.15  & 1.23  & 1.19  & & 1.19  & 1.07  & 1.13  & & 0.83  & 0.74  & 0.85  \\
\cite{heif-ss02}&1.18& 1.22 &       & &       &       &       & &       &       &       \\
\cite{heifss2000}&1.32&1.36 &       & & 1.45  & 1.40  &       & &       &       &       \\
\cite{Cheng-prb62}&\multicolumn{2}{c}{1.21}&1.19& & & &       & &       &       &       \\
\cite{Tinte-prb64}& &       &       & & \multicolumn{2}{c}{1.17}& & &   &       &       \\
\cite{Vand-far114}&\multicolumn{2}{c}{1.26}& & &\multicolumn{2}{c}{1.24}& & &\multicolumn{2}{c}{0.97}& \\
  \hline
  \hline
\end{tabular}
\end{table}
The calculated surface energies of the relaxed surfaces, presented
in Table \ref{surf-en}, were computed using the method described
in Ref. \cite{heif-prb01}. One can see good agreement of the
surface energies obtained by different methods. The energies
calculated for AO- and TiO$_2$-terminated surfaces demonstrate a
small difference, that means both terminations could co-exist.
Nevertheless, the energy computed for TiO$_2$-terminated STO
surface is a little bit larger than that for SrO-termination, in
contrast to BTO and PTO crystals where TiO$_2$-terminated surface
is a little bit energetically more favorable. The three (001)
surfaces of A$^{II}$B$^{IV}$O$_3$ perovskites correspond to ``type
I" stable surfaces revealing the \emph{weak-polarity} due to
partly covalent nature of perovskite  chemical bonding discussed
in the next section.

\subsection{Electronic Charge Redistribution}

We begin discussion of the electronic structure of surfaces with
the analysis of charge redistribution in near-surface planes. The
effective atomic charges (calculated using the standard Mulliken
population analysis) and dipole atomic moments characterizing
atomic deformation along the \emph{Z} axis are presented for all
AO-, TiO$_2$- and asymmetrically terminated surfaces in Tables
\ref{AO-charges}, \ref{TiO-charges}, and \ref{asymm-charges},
respectively. The differences in charge densities in the (001)
planes in ABO$_3$ bulk crystals and on the (001) surfaces are
analyzed in Table \ref{char-densi}.

\begin{table}
\renewcommand{\baselinestretch}{1.0}
\centering
 \caption{
          \small The calculated Mulliken effective charges and dipole moments for the AO termination. Numbers in brackets
          are deviations from bulk values. Bulk charges (in e); STO: Sr = 1.871, Ti = 2.35,
          O = -1.407, BTO: Ba = 1.795, Ti = 2.364, O = -1.386, PTO: Pb = 1.343,
          Ti = 2.335, O = -1.226 \cite{pisk-BS}.
          }
\label{AO-charges}
\renewcommand{\baselinestretch}{1.2}
\small\normalsize\scriptsize
\begin{tabular}{cccccccccc}
   &     &          &          & &          &          & &          &          \\
  \hline
  \hline
 N & Ion & STO      &          & & BTO      &          & & PTO      &          \\
 \cline{3-4} \cline{6-7} \cline{9-10}
   &     & Q, e     & d,       & & Q, e     &  d,      & & Q, e     & d,       \\
   &     &          & e a.u.   & &          &  e a.u.  & &          & e a.u.   \\
\hline
 1 &  A  &  1.845   & -0.2202  & & 1.751    & -0.4634  & & 1.277    & -0.4804  \\
   &     & (-0.026) &          & & (-0.044) &          & & (-0.066) &          \\
   &  O  & -1.524   & -0.0336  & & -1.473   & -0.0532  & & -1.131   & 0.0248   \\
   &     & (-0.117) &          & & (-0.087) &          & & (+0.095) &          \\
 2 &  Ti &  2.363   & 0.0106   & & 2.377    & 0.0070   & & 2.333    & -0.0211  \\
   &     & (+0.013) &          & & (+0.013) &          & & (-0.002) &          \\
   &O$_{2}$&-1.449  & -0.0191  & & -1.417   & 0.0182   & & -1.257   & -0.0062  \\
   &     & (-0.042) &          & & (-0.031) &          & & (-0.031) &          \\
 3 &  A  & 1.875    & -0.0232  & & 1.801    & -0.0433  & & 1.354    & -0.0484  \\
   &     & (+0.004) &          & & (+0.006) &          & & (+0.011) &          \\
   &  O  & -1.429   & 0.0008   & & -1.415   & -0.0084  & & -1.258   & -0.0155  \\
   &     & (-0.022) &          & & (-0.029) &          & & (-0.032) &          \\
 4 &  Ti & 2.336    &    0     & & 2.386    & 0        & & 2.342    & 0        \\
   &     & (-0.014) &          & & (+0.004) &          & & (+0.007) &          \\
   &O$_{2}$&-1.411  &     0    & & -1.392   & 0        & & -1.232   & 0        \\
   &     &  (-0.004)&          & & (-0.006) &          & & (-0.006) &          \\
  \hline
  \hline
\end{tabular}
\end{table}
\begin{table}
\renewcommand{\baselinestretch}{1.0}
\centering \caption{
         \small For the TiO$_{2}$ termination. The same as Table \ref{AO-charges}
         }
\label{TiO-charges}
\renewcommand{\baselinestretch}{1.2}
\small\normalsize\scriptsize
\begin{tabular}{cccccccccc}
   &     &          &          & &          &          & &          &          \\
  \hline
  \hline
 N & Ion & STO      &          & & BTO      &          & & PTO      &          \\
 \cline{3-4} \cline{6-7} \cline{9-10}
   &     & Q, e     & d,       & & Q, e     &  d,      & & Q, e     & d,       \\
   &     &          & e a.u.   & &          &  e a.u.  & &          & e a.u.   \\
\hline
 1 & Ti  & 2.314    & 0.0801   & & 2.304    & 0.0816   & & 2.279    & 0.0962   \\
   &     & (-0.036) &          & & (-0.060) &          & & (-0.056) &          \\
   &O$_{2}$&-1.324  & 0.0418   & & -1.278   & 0.0207   & & -1.182   & -0.0307  \\
   &     & (+0.083) &          & & (+0.108) &          & & (+0.044) &          \\
 2 & A   & 1.851    & 0.0423   & & 1.765    & 0.0949   & & 1.270    & 0.0990   \\
   &     & (-0.020) &          & & (-0.030) &          & & (-0.073) &          \\
   & O   & -1.361   & -0.0436  & & -1.343   & -0.0303  & & -1.166   & -0.0060  \\
   &     & (+0.046) &          & & (+0.043) &          & & (+0.060) &          \\
 3 & Ti  & 2.386    & 0.0139   & & 2.362    & 0.0079   & & 2.332    & 0.0183   \\
   &     & (+0.036) &          & & (-0.002) &          & & (-0.003) &          \\
   &O$_{2}$&-1.389  & -0.0167  & & -1.369   & -0.0108  & & -1.205   & -0.0169  \\
   &     & (+0.018) &          & & (+0.017) &          & & (+0.021) &          \\
 4 &  A  & 1.871    & 0        & & 1.794    & 0        & & 1.337    & 0        \\
   &     & (0.000)  &          & & (-0.001) &          & & (-0.006) &          \\
   &  O  & -1.399   & 0        & & -1.381   & 0        & & -1.219   & 0        \\
   &     & (+0.008) &          & & (+0.005) &          & & (+0.007) &          \\
  \hline
  \hline
\end{tabular}
\end{table}
\begin{table}
\renewcommand{\baselinestretch}{1.0}
\centering \caption{
         \small For the asymmetrical termination. The same as Table \ref{AO-charges}
         }
\label{asymm-charges}
\renewcommand{\baselinestretch}{1.2}
\small\normalsize\scriptsize
\begin{tabular}{cccccccccc}
   &     &          &          & &          &          & &          &          \\
  \hline
  \hline
 N & Ion & STO      &          & & BTO      &          & & PTO      &          \\
 \cline{3-4} \cline{6-7} \cline{9-10}
   &     & Q, e     & d,       & & Q, e     &  d,      & & Q, e     & d,       \\
   &     &          & e a.u.   & &          &  e a.u.  & &          & e a.u.   \\
\hline
 1 & A   & 1.845    & -0.2238  & & 1.751    & -0.4653  & & 1.249    & -0.3055  \\
   &     & (-0.026) &          & & (-0.044) &          & & (-0.094) &          \\
   & O   & -1.520   & -0.0428  & & -1.470   & -0.0548  & & -1.124   & 0.0305   \\
   &     & (-0.113) &          & & (-0.084) &          & & (+0.102) &          \\
 2 & Ti  & 2.361    & 0.0117   & & 2.376    & 0.0054   & & 2.318    & -0.0350  \\
   &     & (+0.011) &          & & (+0.012) &          & & (-0.017) &          \\
   &O$_{2}$&-1.450  & -0.0156  & & -1.417   & 0.0265   & & -1.242   & -0.0493  \\
   &     & (-0.043) &          & & (-0.031) &          & & (-0.016) &          \\
 3 & A   & 1.874    & -0.0225  & & 1.800    & -0.0461  & & 1.345    & 0.0660   \\
   &     & (+0.003) &          & & (+0.005) &          & & (+0.002) &          \\
   & O   & -1.425   & 0.0067   & & -1.414   & -0.0082  & & -1.252   & 0.0062   \\
   &     & (-0.018) &          & & (-0.028) &          & & (-0.026) &          \\
 4 & Ti  & 2.352    & 0.0026   & & 2.366    & 0.0006   & & 2.325    & 0.0025   \\
   &     & (+0.002) &          & & (+0.002) &          & & (-0.010) &          \\
   &O$_{2}$&-1.408  & -0.0006  & & -1.389   & -0.0024  & & -1.223   & -0.0323  \\
   &     & (-0.001) &          & & (-0.003) &          & & (+0.003) &          \\
 5 & A   & 1.870    & -0.0042  & & 1.795    & -0.0074  & & 1.334    & 0.1315   \\
   &     & (-0.001) &          & & (0.000)  &          & & (-0.009) &          \\
   & O   & -1.404   & 0.0102   & & -1.384   & 0.0089   & & -1.222   & 0.0425   \\
   &     & (+0.003) &          & & (+0.002) &          & & (+0.004) &          \\
 6 & Ti  & 2.348    & -0.0089  & & 2.362    & -0.0084  & & 2.325    & -0.0115  \\
   &     & (-0.002) &          & & (-0.002) &          & & (-0.010) &          \\
   &O$_{2}$&-1.383  & 0.0057   & & -1.370   & 0.0103   & & -1.200   & 0.0042   \\
   &     & (+0.024) &          & & (+0.016) &          & & (+0.026) &          \\
 7 & A   & 1.846    & -0.0633  & & 1.765    & -0.0950  & & 1.260    & 0.0019   \\
   &     & (-0.022) &          & & (-0.030) &          & & (-0.083) &          \\
   & O   & -1.365   & 0.0288   & & -1.343   & 0.0318   & & -1.149   & 0.0254   \\
   &     & (+0.042) &          & & (+0.043) &          & & (+0.077) &          \\
 8 & Ti  & 2.291    & -0.0748  & & 2.305    & -0.0823  & & 2.272    & -0.1508  \\
   &     & (-0.059) &          & & (-0.059) &          & & (-0.063) &          \\
   &O$_{2}$&-1.297  & -0.0493  & & -1.278   & -0.0189  & & -1.176   & -0.0048  \\
   &     & (+0.110) &          & & (+0.108) &          & & (+0.050) &          \\
  \hline
  \hline
\end{tabular}
\end{table}
\begin{table}
\renewcommand{\baselinestretch}{1.0}
 \centering
 \caption{
          \small Calculated charge densities in the (001) planes in the bulk perovskites
          (in e, per TiO$_{2}$ or AO unit, data are taken from \cite{pisk-BS}) and
          in four top planes of the AO-, TiO$_{2}$-terminated
           and asymmetrical slabs. Deviations of charge density with
           respect to the bulk are given in brackets.
          }\label{char-densi}
\renewcommand{\baselinestretch}{1.2}
\small\normalsize\scriptsize
\begin{tabular}{cccccc}
               &    &         &           &          &          \\
  \hline
  \hline
  Termination  & N  & Unit    &  STO      &  BTO     & PTO      \\
  \hline
  Bulk         &    & AO      &  0.464    & 0.409    & 0.117    \\
               &    &TiO$_{2}$& -0.464    & -0.409   & -0.117   \\
  \hline
  AO           & 1  & AO      & 0.321     & 0.278    & 0.146    \\
               &    &         & (-0.143)  & (-0.131) & (0.029)  \\
               & 2  &TiO$_{2}$& -0.535    & -0.457   & -0.181   \\
               &    &         & (-0.071)  & (-0.049) & (-0.064) \\
               & 3  & AO      & 0.446     & 0.386    & 0.096    \\
               &    &         & (-0.018)  & (-0.023) & (-0.021) \\
               & 4  &TiO$_{2}$& -0.486    & -0.398   & -0.122   \\
               &    &         & (-0.022)  & (0.010)  & (-0.005) \\
  \hline
 TiO$_{2}$     & 1  &TiO$_{2}$& -0.334    & -0.252   & -0.085   \\
               &    &         & (0.130)   & (0.156)  & (0.032)  \\
               & 2  & AO      & 0.490     & 0.422    & 0.104    \\
               &    &         & (0.026)   & (0.013)  & (-0.013) \\
               & 3  &TiO$_{2}$& -0.392    & -0.376   & -0.078   \\
               &    &         & (0.072)   & (0.032)  & (0.039)  \\
               & 4  & AO      & 0.472     & 0.413    & 0.118    \\
               &    &         & (0.008)   & (0.004)  & (0.001)  \\
  \hline
 Asymmetrical  & 1  & AO      & 0.325     & 0.281    & 0.125    \\
               &    &         & (-0.139)  & (-0.128) & (0.008)  \\
               & 2  &TiO$_{2}$& -0.539    & -0.458   & -0.166   \\
               &    &         & (-0.075)  & (-0.050) & (-0.049) \\
               & 3  & AO      & 0.449     & 0.386    & 0.093    \\
               &    &         & (-0.015)  & (-0.023) & (-0.024) \\
               & 4  &TiO$_{2}$& -0.464    & -0.421   & -0.121   \\
               &    &         & (0.000)   & (-0.004) & (-0.004) \\
               & 5  & AO      & 0.466     & 0.411    & 0.112    \\
               &    &         & (0.002)   & (0.002)  & (-0.005) \\
               & 6  &TiO$_{2}$& -0.418    & -0.378   & -0.075   \\
               &    &         & (0.046)   & (0.030)  & (0.042)  \\
               & 7  & AO      & 0.481     & 0.422    & 0.111    \\
               &    &         & (0.017)   & (0.013)  & (-0.006) \\
               & 8  &TiO$_{2}$& -0.303    & -0.251   & -0.080   \\
               &    &         & (0.161)   & (0.157)  & (0.037)  \\
  \hline
  \hline
\end{tabular}
\end{table}

First of all, note that the effective charges of Sr and Ba are
close to the +2e formal charges, whereas that of Pb is
considerably smaller (~2$e$). Ti and O charges are also much
smaller than formal charges, similarly to the bulk \cite{pisk-BS}
which results from the Ti-O covalent bounding. The AO-terminated
surfaces of STO and BTO show similar behavior. The charges of top
layer cations are smaller with respect to the relevant bulk
charges, the oxygens attract an additional electron charge and
become more negative. The titanium ions in the second layer
demonstrate the slight increase of their charges, their oxygens
again became more negative due to addition electron charge
transfer. Changes in atomic charges in deeper layers become very
small and practically equal zero in the center of the slabs.
Unlike STO and BTO, the surface oxygen on PbO-terminated surface
becomes more positive, Ti in the second layer shows practically no
changes in the effective charges.

The charge redistribution in the TiO$_2$-terminated surfaces
(Table \ref{TiO-charges}) of all three perovskites demonstrates
quite similar behavior. All cations in both topmost layers
demonstrate charge reduction. For surface Ti it is a little bit
larger then for A ions in subsurface layer. Changes of charges for
ions in the asymmetrically terminated slabs (Table
\ref{asymm-charges}) are practically the same as for the
symmetrically terminated AO- and TiO$_2$- slab as it should be
when slabs are thick enough and the two surfaces do not iteract.

The calculated effective charges of ions in partly-covalent
materials usually are far from the formal ionic charges due to
electron density redistribution caused by the covalency effect,
what is confirmed by our calculations (see also heading to Table
\ref{AO-charges}). As a result, the AO and TiO$_2$ (001) planes in
the bulk perovskites turn out to be \emph{charged} with charge
density per unit cell: $\sigma _B (AO)=-\sigma _B (TiO_2)$ (B
means ``Bulk", Table \ref{char-densi}). A half of this charge
density comes from TiO$_2$ planes to each of the two neighboring
AO planes. If the formal ionic charges (+4e, -2e, +2e) took place,
the charge densities of the (001) planes would be zero and we deal
with the neutral, ``type-I surfaces" (according to the generally-accepted
classification by Tasker \cite{tasker-JPC-SSP12}). In reality, the
charge redistribution makes the ABO$_3$(001) surfaces to be polar
with the dipole moment perpendicular to the surface or ``type-III
polar surfaces", which are potentially unstable. In case of
symmetrically terminated slabs this dipole moment disappears due
to symmetry, but the case of asymmetrical slabs is quite of
interest. One of possibilities to stabilize the surface is to
create surface defects \cite{heif-ss02}. Another option is to
add the compensating charge density to the topmost layers of
surfaces. The charge density changes for all perovskite surfaces
with respect to the bulk magnitudes are summarized in Table
\ref{char-densi}. It is well seen, as a result of self-consistent
calculations, the additional charge densities indeed exist, these
are localized mainly on two upper layers whereas the central
layers practically retain the bulk charge density. Such the
electron charge density redistribution accompanied by atomic
displacements on the surfaces allows to compensate the surface
dipole moment even for the asymmetrically terminated slabs. Indeed,
as one can see, the sum of changes in charge densities for three
topmost layers of the asymmetrically terminated slab (on both
sides) of all three perovskites approximately equals half the
charge density of the corresponding bulk plane, which is necessary
condition to remove the macroscopic dipole moment
 \cite{tasker-JPC-SSP12}. These results are in a
line with ideas of ``weak polarity"
\cite{noguerab,noguera-SS365}.

The atomic \emph{dipole moments} characterize the deformation and
polarization along the $z$-axis perpendicular to the surface
\cite{CRman}. These are presented in Tables \ref{AO-charges},
\ref{TiO-charges}, \ref{asymm-charges} for AO-, TiO$_2$- and
asymmetrically terminated surfaces, respectively. On the
AO-terminated surfaces of all three perovskites the cations have
the negative dipole moments, directed inwards, to the slab center.
The dipole moments of surface Sr, Ba and Pb cations on
AO-terminated surfaces are surprisingly large, several times
larger than those of other ions, including the TiO$_2$-terminated
surfaces. Oxygens of the AO-terminated surfaces of STO and BTO
demonstrate small negative dipole moments whereas for the PTO the
oxygen dipole moment is positive. The polarization of second and
third layers is pretty small.

On the TiO$_2$ terminated surfaces polarization of cations has a
positive sign, as well as for the oxygens in STO and BTO. In
contrast, oxygens on PbO surface have negative dipole moment.
Cations of subsurface layers for all three perovskites reveal the
positive dipole moments whereas those of oxygens are negative. The
asymmetrically terminated slabs, actually, reproduce the
polarization picture obtained for the two relevant symmetrically
terminated slabs.

\begin{table}
\renewcommand{\baselinestretch}{1.0}
 \centering
 \caption{
          \small The bond populations for the AO termination (in me, m$=$milli). Negative
          bond population means atomic repulsion. The corresponding
          bond populations for the bulk perovskites: Ti-O bond: (STO)
          88; (BTO) 100; (PTO) 98; Pb-O bond: 16.
          }
 \label{AO-bond}
\renewcommand{\baselinestretch}{1.2}
\small\normalsize\scriptsize
\begin{tabular}{ccccccccc}
   &        &          &        &        &        &        & &        \\
  \hline
  \hline
      &  STO    &      & & BTO     &       & & PTO     &        \\
 \cline{2-3} \cline{5-6} \cline{8-9}
Atom A& Atom B  &      & & Atom B  &       & & Atom B  &        \\
  \hline
 O(1) & O(1)    & 4    & & O(1)    & 2     & & O(1)    &  0     \\
      & Sr(1)   & -6   & & Ba(1)   & -30   & & Pb(1)   &  54    \\
      & Ti(2)   & 72   & & Ti(2)   & 80    & & Ti(2)   &  102   \\
      & O(2)    & -54  & & O(2)    & -58   & & O(2)    &  -74   \\
 O(2) & Sr(1)   & -30  & & Ba(1)   & -56   & & Pb(1)   &  52    \\
      & O(2)    & -46  & & O(2)    & -38   & & O(2)    &  -60   \\
      & Ti(2)   & 78   & & Ti(2)   & 88    & & Ti(2)   &  80    \\
      & Sr(3)   & -10  & & Ba(3)   & -30   & & Pb(3)   &  6     \\
      & O(3)    & -48  & & O(3)    & -34   & & O(3)    &  -42   \\
 O(3) & Ti(2)   & 86   & & Ti(2)   & 90    & & Ti(2)   &  72    \\
      & O(3)    & -8   & & O(3)    & -6    & & O(3)    &  -8    \\
      & Sr(3)   & -12  & & Ba(3)   & -36   & & Pb(3)   &  24    \\
      & Ti(4)   & 84   & & Ti(4)   & 98    & & Ti(4)   &  96    \\
      & O(4)    & -46  & & O(4)    & -38   & & O(4)    &  -54   \\
 O(4) & Sr(3)   & -10  & & Ba(3)   & -34   & & Pb(3)   &  24    \\
      & O(4)    & -8   & & O(4)    & -6    & & O(4)    &  -8    \\
      & Ti(4)   & 86   & & Ti(4)   & 98    & & Ti(4)   &  94    \\
  \hline
  \hline
\end{tabular}
\end{table}
\begin{table}
\renewcommand{\baselinestretch}{1.0}
 \centering
 \caption{
          \small The same as Table \ref{AO-bond} for the TiO$_{2}$ termination.
          }
 \label{TiO-bond}
\renewcommand{\baselinestretch}{1.2}
\small\normalsize\scriptsize
\begin{tabular}{ccccccccc}
   &        &          &        &        &        &        & &        \\
  \hline
  \hline
      &  STO    &      & & BTO     &       & & PTO     &        \\
 \cline{2-3} \cline{5-6} \cline{8-9}
Atom A& Atom B  &      & & Atom B  &       & & Atom B  &        \\
  \hline
 O(1) & O(1)    & -30  & & O(1)    & -24   & & O(1)    & -34    \\
      & Ti(1)   & 114  & & Ti(1)   & 126   & & Ti(1)   & 114    \\
      & Sr(2)   & -14  & & Ba(2)   & -38   & & Pb(2)   & 42     \\
      & O(2)    & -28  & & O(2)    & -20   & & O(2)    & -42    \\
 O(2) & Ti(1)   & 142  & & Ti(1)   & 140   & & Ti(1)   & 162    \\
      & O(2)    & 2    & & O(2)    & 2     & & O(2)    & 0      \\
      & Sr(2)   & -8   & & Ba(2)   & -30   & & Pb(2)   & 8      \\
      & Ti(3)   & 72   & & Ti(3)   & 90    & & Ti(3)   & 80     \\
      & O(3)    & -36  & & O(3)    & -32   & & O(3)    & -36    \\
 O(3) & Sr(2)   & -4   & & Ba(2)   & -24   & & Pb(2)   & 14     \\
      & O(3)    & -42  & & O(3)    & -36   & & O(3)    & -44    \\
      & Ti(3)   & 94   & & Ti(3)   & 106   & & Ti(3)   & 110    \\
      & Sr(4)   & -10  & & Ba(4)   & -34   & & Pb(4)   & 18     \\
      & O(4)    & -42  & & O(4)    & -36   & & O(4)    & -44    \\
 O(4) & Ti(3)   & 92   & & Ti(3)   & 102   & & Ti(3)   & 106    \\
      & O(4)    & 2    & & O(4)    & 2     & & O(4)    & 2      \\
      & Sr(4)   & -10  & & Ba(4)   & -34   & & Pb(4)   & 14     \\
  \hline
  \hline
\end{tabular}
\end{table}
\begin{table}
\renewcommand{\baselinestretch}{1.0}
 \centering
 \caption{
          \small The same as Table \ref{AO-bond} for the asymmetrical termination.
          }
 \label{asymm-bond}
\renewcommand{\baselinestretch}{1.2}
\small\normalsize\scriptsize
\begin{tabular}{ccccccccc}
   &        &          &        &        &        &        & &        \\
  \hline
  \hline
      &  STO    &      & & BTO     &       & & PTO     &        \\
 \cline{2-3} \cline{5-6} \cline{8-9}
Atom A& Atom B  &      & & Atom B  &       & & Atom B  &        \\
  \hline
 O(1) & O(1)    & 4    & & O(1)    & 2     & & O(1)    & 0      \\
      & Sr(1)   & -6   & & Ba(1)   & -30   & & Pb(1)   & 58     \\
      & Ti(2)   & 70   & & Ti(2)   & 82    & & Ti(2)   & 104    \\
      & O(2)    & -56  & & O(2)    & -60   & & O(2)    & -70    \\
 O(2) & Sr(1)   & -32  & & Ba(1)   & -58   & & Pb(1)   & 48     \\
      & O(2)    & -46  & & O(2)    & -38   & & O(2)    & -58    \\
      & Ti(2)   & 76   & & Ti(2)   & 88    & & Ti(2)   & 88     \\
      & Sr(3)   & -10  & & Ba(3)   & -30   & & Pb(3)   & 6      \\
      & O(3)    & -48  & & O(3)    & -32   & & O(3)    & -44    \\
 O(3) & Ti(2)   & 86   & & Ti(2)   & 90    & & Ti(2)   & 82     \\
      & O(3)    & 2    & & O(3)    & -6    & & O(3)    & 0      \\
      & Sr(3)   & -12  & & Ba(3)   & -36   & & Pb(3)   & 26     \\
      & Ti(4)   & 86   & & Ti(4)   & 98    & & Ti(4)   & 98     \\
      & O(4)    & -46  & & O(4)    & -38   & & O(4)    & -52    \\
 O(4) & Sr(3)   & -10  & & Ba(3)   & -34   & & Pb(3)   & 14     \\
      & O(4)    & -8   & & O(4)    & 2     & & O(4)    & -8     \\
      & Ti(4)   & 88   & & Ti(4)   & 98    & & Ti(4)   & 100    \\
      & Sr(5)   & -10  & & Ba(5)   & -34   & & Pb(5)   & 14     \\
      & O(5)    & -44  & & O(5)    & -38   & & O(5)    & -50    \\
 O(5) & Ti(4)   & 88   & & Ti(4)   & 102   & & Ti(4)   & 102    \\
      & O(5)    & 2    & & O(5)    & -6    & & O(5)    & 0      \\
      & Sr(5)   & -10  & & Ba(5)   & -34   & & Pb(5)   & 18     \\
      & Ti(6)   & 88   & & Ti(6)   & 102   & & Ti(6)   & 102    \\
      & O(6)    & -44  & & O(6)    & -36   & & O(6)    & -50    \\
 O(6) & Sr(5)   & -10  & & Ba(5)   & -34   & & Pb(5)   & 24     \\
      & O(6)    & -42  & & O(6)    & -36   & & O(6)    & -44    \\
      & Ti(6)   & 96   & & Ti(6)   & 104   & & Ti(6)   & 110    \\
      & Sr(7)   & -4   & & Ba(7)   & -24   & & Pb(7)   & 14     \\
      & O(7)    & -36  & & O(7)    & -32   & & O(7)    & -34    \\
 O(7) & Ti(6)   & 80   & & Ti(6)   & 90    & & Ti(6)   & 84     \\
      & O(7)    & 2    & & O(7)    & 2     & & O(7)    & 0      \\
      & Sr(7)   & -8   & & Ba(7)   & -30   & & Pb(7)   & 8      \\
      & Ti(8)   & 132  & & Ti(8)   & 142   & & Ti(8)   & 160    \\
      & O(8)    & -28  & & O(8)    & -20   & & O(8)    & -40    \\
 O(8) & Sr(7)   & -14  & & Ba(7)   & -38   & & Pb(7)   & 42     \\
      & Ti(8)   & 118  & & Ti(8)   & 126   & & Ti(8)   & 116    \\
      & O(8)    & -32  & & O(8)    & -24   & & O(8)    & -32    \\
  \hline
  \hline
\end{tabular}
\end{table}
The Mulliken \emph{bond populations} between atoms in surface
layers (which arise due to covalency) are presented for the AO-,
TiO$_2$- and asymmetrically terminated slabs in Tables
\ref{AO-bond}, \ref{TiO-bond} and \ref{asymm-bond}, respectively.
The largest effect is observed for PTO crystal: the population of
the Pb-O bond on the top layer increases a factor of three, as
compared to the bulk. The partly covalent nature of Pb-O bond in
lead titanate crystal due to hybridization of Pb 6\emph{s} AO
state with the O 2\emph{p} AO is already pronounced in the bulk
\cite{pisk-BS}, but due to bond shortening (due to surface
relaxation) its covalency is increased. This effect is also
observed on the TiO$_2$ surface, for the bond population between
surface O and Pb in the second plane. Unlike PTO, in STO and BTO
there is no indication on the Sr-, Ba- bonding with O atoms.
Unlike Pb, these cations have effective charges close to formal
charge +2e (Table \ref{AO-charges}). The Ti-O bonds of all three
perovskites on the TiO$_2$ terminated surfaces increase their
covalency due to bond shortening (caused by surface relaxation)
and breaking O surface bonds (Table \ref{TiO-bond}). In line with
Ref. \cite{heif-ss02,heifss2000}, we observe $\approx50\%$
increase of the Ti-O bond covalency on the (001) surface. The
asymmetrically terminated slabs demonstrate practically the same
bond populations as discussed above AO- and TiO$_2$-terminated
slabs.

\setlength{\fboxsep}{1pt}
\begin{figure}[htbp]
\def\baselinestretch{1.0}
\setlinespacing{1.0}
  \begin{center}
    \epsfig{file=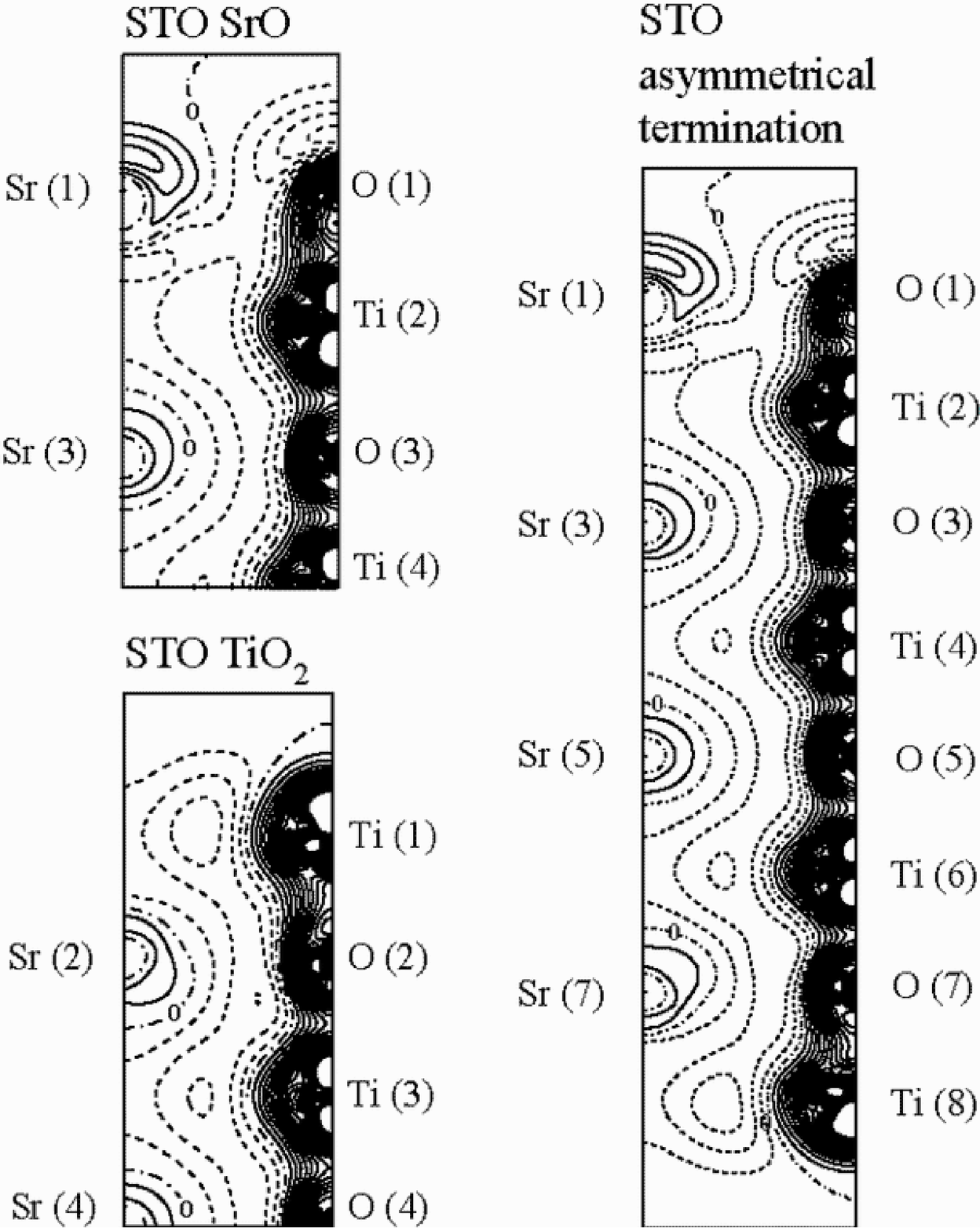,angle=0,width=9cm}
    \caption{
\small The difference electron density maps in the cross section
perpendicular to the (001) surface ((110) plane) with the AO-,
TiO$_2$ and asymmetrical terminations. Isodensity curves are drawn
from -0.05 to +0.05 e a.u.$^{-3}$ with an increment of 0.0025 e
a.u.$^{-3}$.
      }
    \label{surf-ECHD-STO}
  \end{center}
\end{figure}
\setlength{\fboxsep}{1pt}
\begin{figure}[htbp]
\def\baselinestretch{1.0}
\setlinespacing{1.0}
  \begin{center}
    \epsfig{file=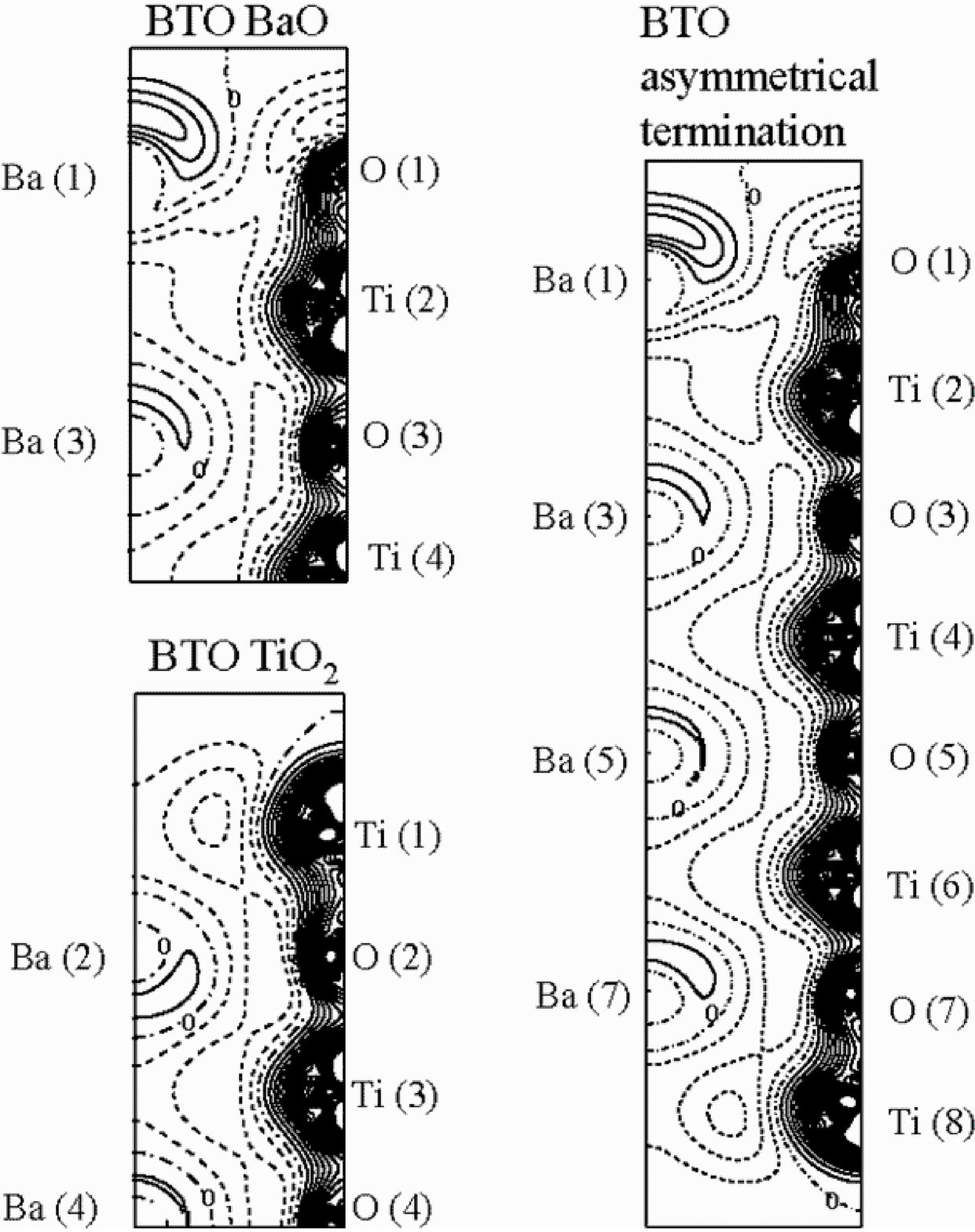,angle=0,width=9cm}
    \caption{
\small The BTO difference electron density maps. The same as Fig.
\ref{surf-ECHD-STO}.
      }
    \label{surf-ECHD-BTO}
  \end{center}
\end{figure}
\setlength{\fboxsep}{1pt}
\begin{figure}[htbp]
\def\baselinestretch{1.0}
\setlinespacing{1.0}
  \begin{center}
    \epsfig{file=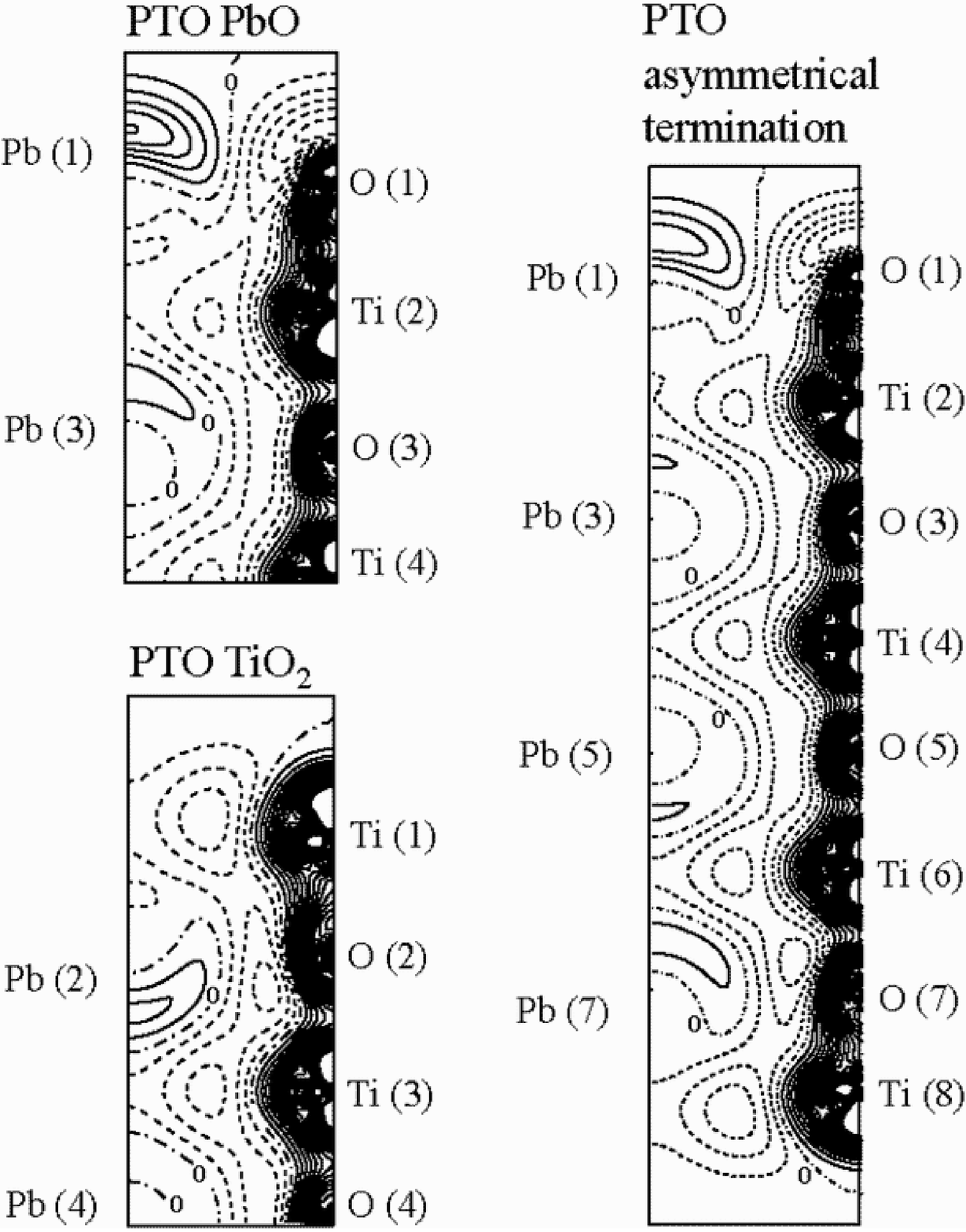,angle=0,width=9cm}
    \caption{
\small The PTO difference electron density maps. The same as Fig.
\ref{surf-ECHD-STO}.
      }
    \label{surf-ECHD-PTO}
  \end{center}
\end{figure}
The \emph{difference electron density maps} calculated with
respect to the superposition density of spherical A$^{2+}$,
Ti$^{4+}$ and O$^{2-}$ ions for STO, BTO and PTO surfaces are
presented in Fig. \ref{surf-ECHD-STO}, \ref{surf-ECHD-BTO} and
\ref{surf-ECHD-PTO}, respectively. These maps demonstrate
considerable electron charge density redistribution near the
perovskite surfaces and are entirely consistent with the
above-discussed Mulliken charges and bond population analysis. For
all three perovskites the excess of electron density (the solid
isodensity curves) is observed for the Ti-O bonds, which
corresponds to the bond covalency. For all terminations nearest to
the surface Ti-O bond becomes stronger, but the next nearest bond
becomes weaker. The A cations on the AO-terminated surfaces
demonstrate considerable polarization, as it was predicted above
from dipole moment calculations (Tables \ref{AO-charges},
\ref{TiO-charges} and \ref{asymm-charges}). Nevertheless, the
electron density maps demonstrate no clear trace of the covalent
bonding (zero dot-dashed curves in area between A cations and Ti-O
pairs) between A cations and oxygen, even for PTO, despite the
Pb-O bond population (Table \ref{AO-bond}) was calculated as 54
me. This means, in reality the covalent contribution in Pb-O bond
in PTO is quite weak and plays negligible role in the covalency of
a whole PTO crystal. Note that the calculated Pb-O bond population
considerably depends on the Mulliken population analysis and the
Pb atomic BS \cite{pisk-BS}, whose the most diffuse exponent
(0.142 bohr$^{-2}$) mainly contributes into the increase of
covalency. Indeed, recently suggested new approach for the
population analysis using minimal valence basis of Wannier Type
Atomic Orbitals (WTAO) \cite{evar-WTAO} (which are directly
connected to the electronic band structure) predicts practicaly
ionic Pb charge $+1.99e$ for the bulk PTO crystal. Thus, our
calculated Pb-O bond population can be interpreted as an increase
of the electron attraction between Pb and O ions rather a real
covalency.

\subsection{Density of States and Band Structures}

\begin{table}
\renewcommand{\baselinestretch}{1.0}
 \centering
 \caption{
          \small The calculated optical gap (in eV) for the bulk \cite{pisk-BS}
          and surface-terminated perovskites. The numbers in brackets are from
          Ref. \cite{heif-ss02} for STO and
           Ref. \cite{lasaro} for PTO.
          Both these calculations had no $d$-orbitals on O, and the later one is done with B3LYP functional. The last row
          contains experimental data.
          }
 \label{gap}
\renewcommand{\baselinestretch}{1.2}
\small\normalsize\scriptsize
\begin{tabular}{ccccc@{\hspace{1mm}}ccccc@{\hspace{1mm}}ccccc}
               &       &       &       &       & &       &       &       &       & &       &       &       &       \\
  \hline
  \hline
               & \multicolumn{4}{c}{STO}& &\multicolumn{4}{c}{BTO}& &\multicolumn{4}{c}{PTO}\\
  \cline{2-5} \cline{7-10} \cline{12-15}
               & bulk  & SrO   &TiO$_{2}$& asm& &bulk & BaO   &TiO$_{2}$& asm& &bulk & PbO   &TiO$_{2}$& asm \\
  \hline
  \multicolumn{15}{c}{Direct gap}\\
  \hline
$\Gamma$-$\Gamma$& 3.96& 3.72  & 3.95  & 3.03  & & 3.55  & 3.49  & 2.96  & 2.73  & & 4.32  & 3.58  & 3.18  & 3.08  \\
               &(4.43) &(4.12) &(3.78) &       & &       &       &       &       & &(4.43) &(3.61) &(3.77) &       \\
  X-X          & 4.53  & 4.37  & 4.04  & 4.09  & & 4.39  & 4.22  & 3.63  & 3.72  & & 3.02  & 3.79  & 3.10  & 3.28  \\
               &(5.08) &(4.70) &(4.38)&       & &       &       &       &       & &(3.21) &(3.82) &(3.12) &       \\
  M-M          & 5.70  & 5.62  & 5.17  & 4.66  & & 5.39  & 5.40  & 4.17  & 4.17  & & 5.55  & 5.37  & 5.01  & 4.88  \\
               &(6.45) &(5.94) &(5.04) &       & &       &       &       &       & &(5.80) &(6.02) &(4.89) &       \\
  R-R          & 6.47  &       &       &       & & 6.12  &       &       &       & & 5.98  &       &       &       \\
               &(7.18) &       &       &       & &       &       &       &       & &       &       &       &       \\
  \hline
  \multicolumn{15}{c}{Indirect gap}\\
  \hline
X-$\Gamma$     & 4.39  & 3.55  & 3.92  & 3.41  & & 4.20  & 3.49  & 3.41  & 3.18  & & 2.87  & 2.96  & 2.98  & 2.78  \\
               &       &       &       &       & &       &       &       &       & &(3.18) &(3.03) &(3.12) &       \\
M-$\Gamma$     & 3.71  & 3.30  & 3.17  & 2.31  & & 3.60  & 3.32  & 2.33  & 2.10  & & 3.66  & 3.55  & 3.19  & 2.96  \\
               &(4.23) &(3.71) &(3.09) &       & &       &       &       &       & &(3.85) &(4.05) &(2.99) &       \\
R-$\Gamma$     & 3.63  &       &       &       & & 3.50  &       &       &       & & 3.66  &       &       &       \\
               &(4.16) &       &       &       & &       &       &       &       & &       &       &       &       \\
  \hline
  \multicolumn{15}{c}{LDA-DFT PF (Ref. \cite{Vand-far114})}\\
  \hline
              &1.85&1.86&1.13& & &1.79&1.80    &0.84   &       & &1.54   &1.53   &1.61   &       \\
  \hline
  \multicolumn{15}{c}{Experiment}\\
  \hline
               &\multicolumn{4}{c}{3.75 - direct gap}& & \multicolumn{4}{c}{3.2} & & \multicolumn{4}{c}{3.4}       \\
               &\multicolumn{4}{c}{3.25 - indirect gap}& &\multicolumn{4}{c}{Ref. \cite{wemple}}& &\multicolumn{4}{c}{Ref. \cite{peng-chang}}\\
               &\multicolumn{4}{c}{Ref. \cite{benthem}}& & &     &       &       & &       &       &       &       \\
  \hline
  \hline
\end{tabular}
\end{table}
\begin{table}
\renewcommand{\baselinestretch}{1.0}
 \centering
 \caption{
          \small The calculated  positions of the valence band top $\varepsilon_v$ and of the conduction band bottom $\varepsilon_c$
          (in eV) for relaxed and unrelaxed perovskite surface structures. The
          values in brackets are results from Ref. \cite{lasaro} for PTO.
          The conduction band bottom is in $\Gamma$-point. The valence band
          top lies at the M-point for STO and BTO,
          but in X-point for PTO.
          In the Ref. \cite{lasaro} the valence band top at the TiO$_{2}$-terminated surface is in M-point.
          }
 \label{rel-unrel-EcEvgap}
\renewcommand{\baselinestretch}{1.2}
\small\normalsize\scriptsize
\begin{tabular}{ccccccccc}
                      &       &              & &              &              & &              &              \\
  \hline
  \hline
                      & \multicolumn{2}{c}{STO}& &\multicolumn{2}{c}{BTO}    & &\multicolumn{2}{c}{PTO}      \\
  \cline{2-3} \cline{5-6} \cline{8-9}
                      & SrO   &TiO$_{2}$     & & BaO          &TiO$_{2}$     & & PbO          &TiO$_{2}$     \\
  \hline
\multicolumn{9}{c}{Unrelaxed structure}\\
  \hline
$\varepsilon_c$       & 1.21  &  -2.59       & &   0.51       &  -2.86       & &   -1.67      & -2.39        \\
$\varepsilon_v$       & -2.48 &  -5.22       & &   -3.01      &  -5.39       & &   -4.59      & -5.14        \\
  \hline
\multicolumn{9}{c}{Relaxed structure}\\
  \hline
$\varepsilon_c$       & -0.50 & -2.78        & &      0.40    & -4.00        & &        -2.11 (-2.13) & -3.04 (-2.08)       \\
$\varepsilon_v$       & -3.80 & -5.95        & &      -2.92   & -6.33        & &        -5.07 (-5.16) & -6.02 (5.07)        \\
  \hline
  \hline
\end{tabular}
\end{table}
The calculated \emph{band structures} for STO and BTO bulk and
surfaces (Fig. \ref{sto-band} and \ref{bto-band}) are quite
similar. The band structure for the $bulk$ perovskites was
calculated using a unit cell which is fourfold extended along the
$z$-axis. Such a supercell is similar to the eight-layer ``slab"
periodically repeated in 3D space and provides also the most
natural comparison with the surface band structures. In the bulk
band structure calculations the bands are plotted using the
$\Gamma$-X-M-$\Gamma$ directions of the typical ``surface"
Brillouin zone (which is simply a square in cubic crystals with
the standard high symmetry points: $\Gamma$ in the center, M the
corner and X the center of square edge). The upper valence bands
(VB) for the STO and BTO bulk are quite flat with the top in M
point and perfectly flat fragment between M and X points. The main
contribution into the upper VB comes from O 2$p_x$ and 2$p_y$ as
it is well seen from the calculated density of states (DOS)
projected onto the corresponding atomic orbitals (AOs) (see Fig.
\ref{sto-1x1x4-dos} and \ref{bto-1x1x4-dos}). The bottom of lowest
conduction band (CB) lies at the $\Gamma$ point with quite flat
fragment between the $\Gamma$ and X points and consist of Ti 3$d$
threefold degenerated T$_{2_{g}}$ level. The optical band gaps for
surfaces and bulk of all three perovskites as calculated by means
of the DFT-B3PW are presented in Table \ref{gap}. One can see a
good agreement with experiment. We should stress here remarkable
agreement of the B3PW bulk gap with the experiment for STO (3.6 eV
\emph{vs} 3.3 eV). This is in a sharp contrast with the typical HF
overestimate of the gap and DFT underestimate (\emph{e.g.} 1.8 eV
for STO, BTO \cite{vand-prb56,vand-ss418}).

\begin{figure}[htbp]
\def\baselinestretch{1.0}
\setlinespacing{1.0}
  \begin{center}
    \epsfig{file=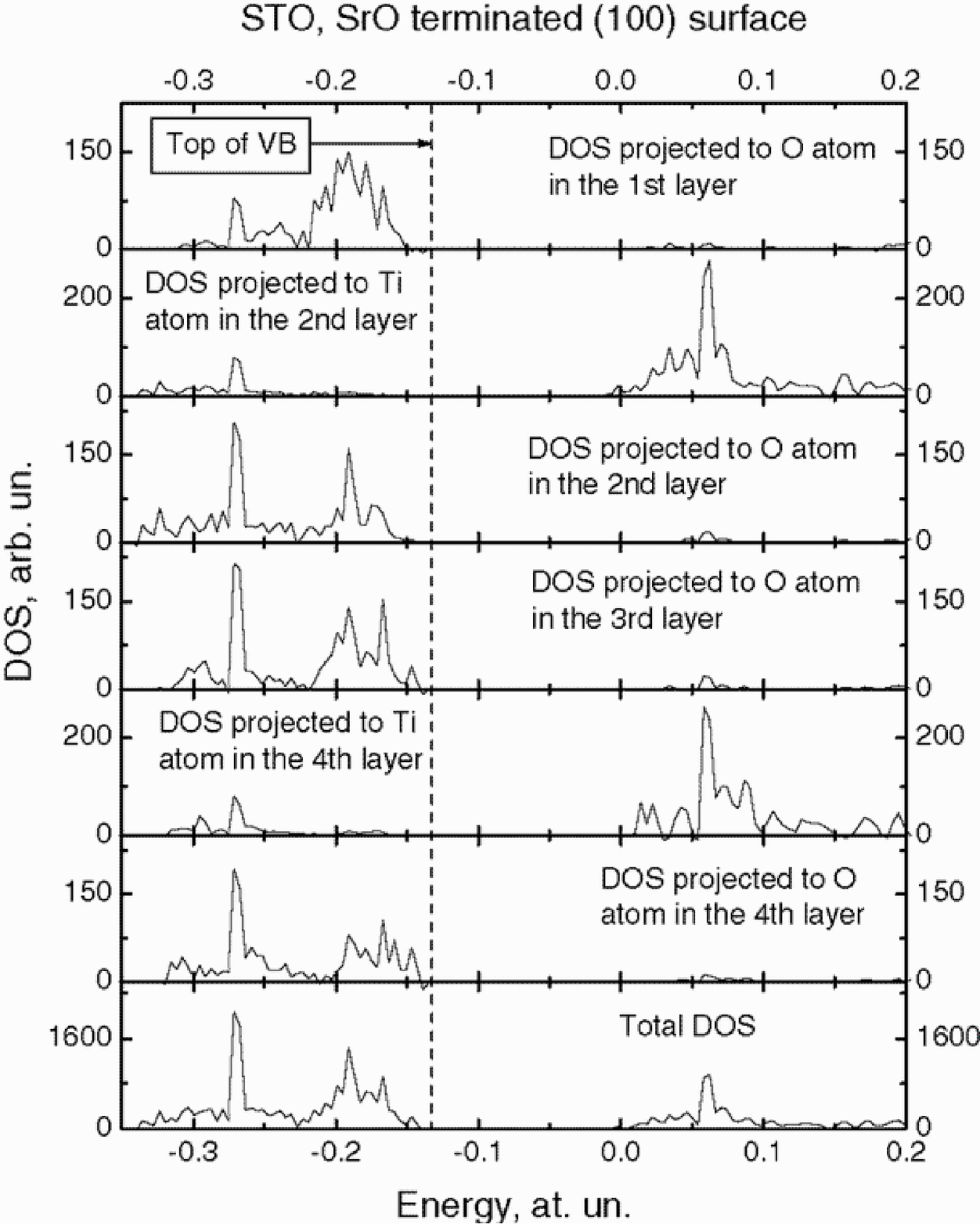,angle=0,width=15cm}
    \caption{
\small Total and projected DOS for the SrO-terminated surface.
      }
    \label{sto-sro-dos}
  \end{center}
\end{figure}

The band structure for the SrO-terminated surface demonstrates
practically the same flat the upper VB as in the bulk STO, with
the top of VB at the M point and the bottom of CB at the $\Gamma$.
The optical band gap for the SrO-terminated surface becomes
smaller with respect to the band gap of the bulk STO. The narrow
indirect gap between the $\Gamma$ and M points is 3.3 eV, whereas
the narrowest gap in the bulk is 3.63 eV, \emph{i.e.} surface gap
is reduced by ~0.3 eV (see Table \ref{gap} for details). Analysis
of the Density Of States (DOS)  calculated for the SrO-terminated surface (Fig.
\ref{sto-sro-dos}) demonstrates no contribution of the surface O
$2p$ states into the top of VB which mainly consists of $2p$ AOs
of the oxygens from the \emph{central} plane. The main contribution into the CB
bottom comes from the Ti $3d$ which are in the second layer.

\setlength{\fboxsep}{1pt}
\begin{figure}[htbp]
\def\baselinestretch{1.0}
\setlinespacing{1.0}
\begin{center}
\epsfig{file=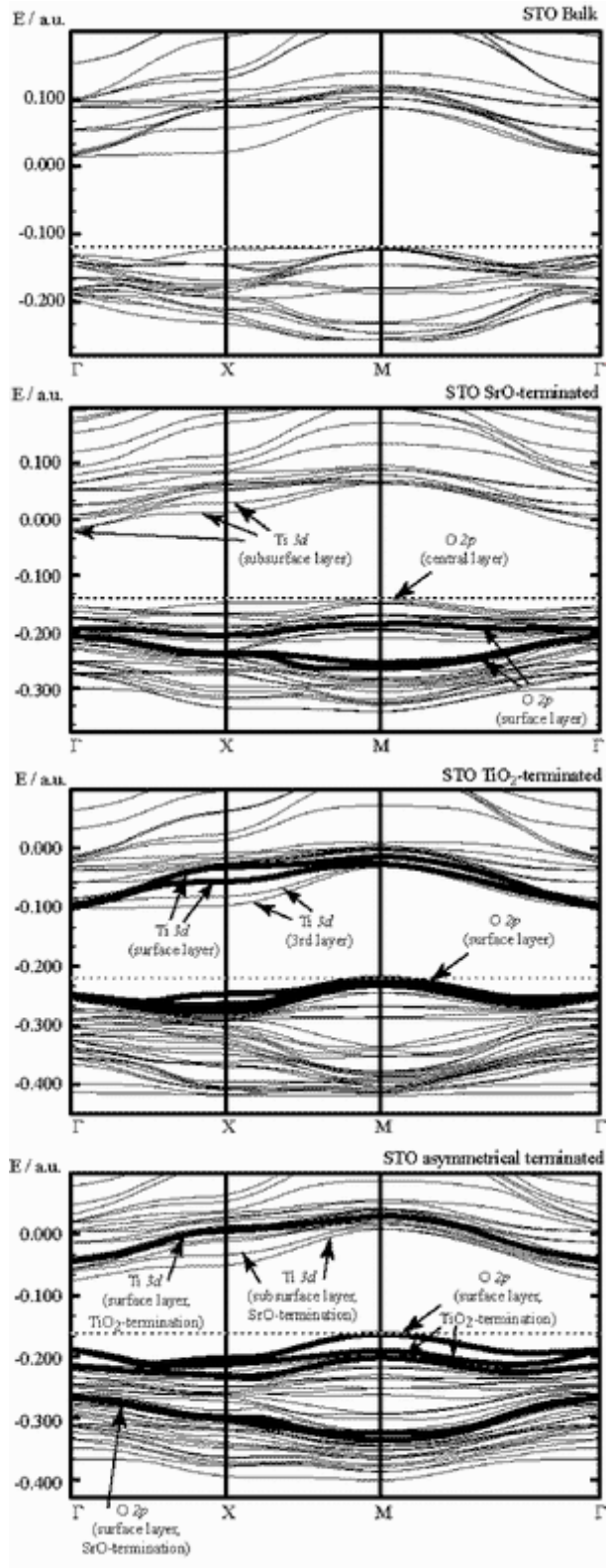,angle=0,width=8cm}
\caption {
         \small The calculated electronic band structure for STO bulk and
         surfaces.
         }
\label{sto-band}
\end{center}
\end{figure}
\setlength{\fboxsep}{1pt}
\begin{figure}[htbp]
\def\baselinestretch{1.0}
\setlinespacing{1.0}
\begin{center}
\epsfig{file=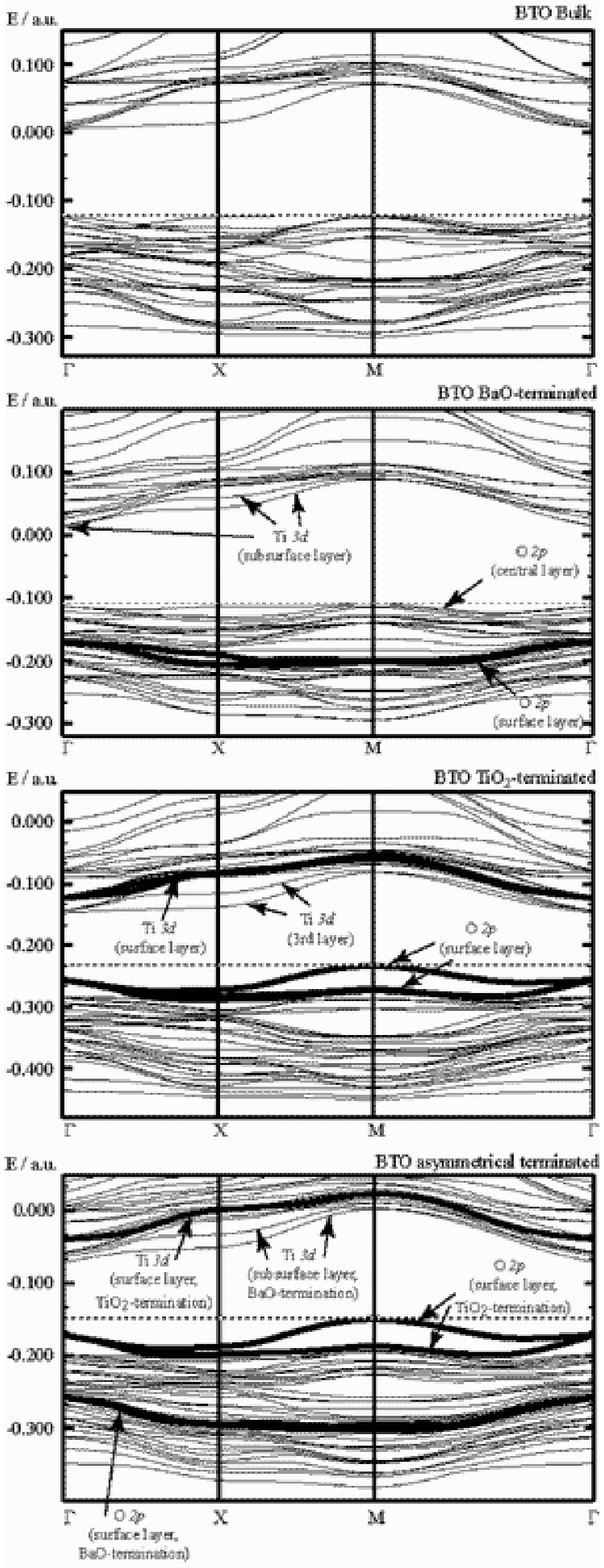,angle=0,width=8cm}
\caption {
         \small The calculated electronic band structure for BTO bulk and
         surfaces.
         }
\label{bto-band}
\end{center}
\end{figure}
\setlength{\fboxsep}{1pt}
\begin{figure}[htbp]
\def\baselinestretch{1.0}
\setlinespacing{1.0}
\begin{center}
\epsfig{file=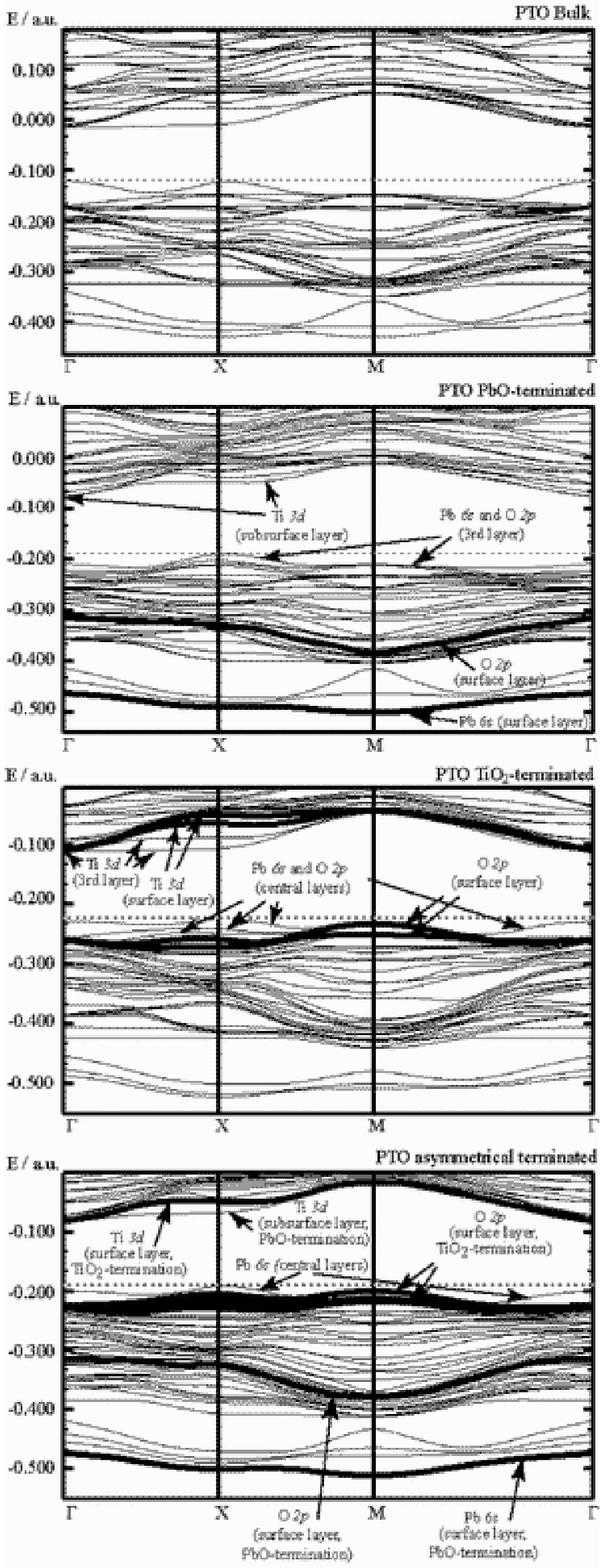,angle=0,width=8cm}
\caption {
         \small The calculated electronic band structure for PTO bulk and
         surfaces.
         }
\label{pto-band}
\end{center}
\end{figure}
\setlength{\fboxsep}{1pt}
\begin{figure}[htbp]
\def\baselinestretch{1.0}
\setlinespacing{1.0}
  \begin{center}
    \epsfig{file=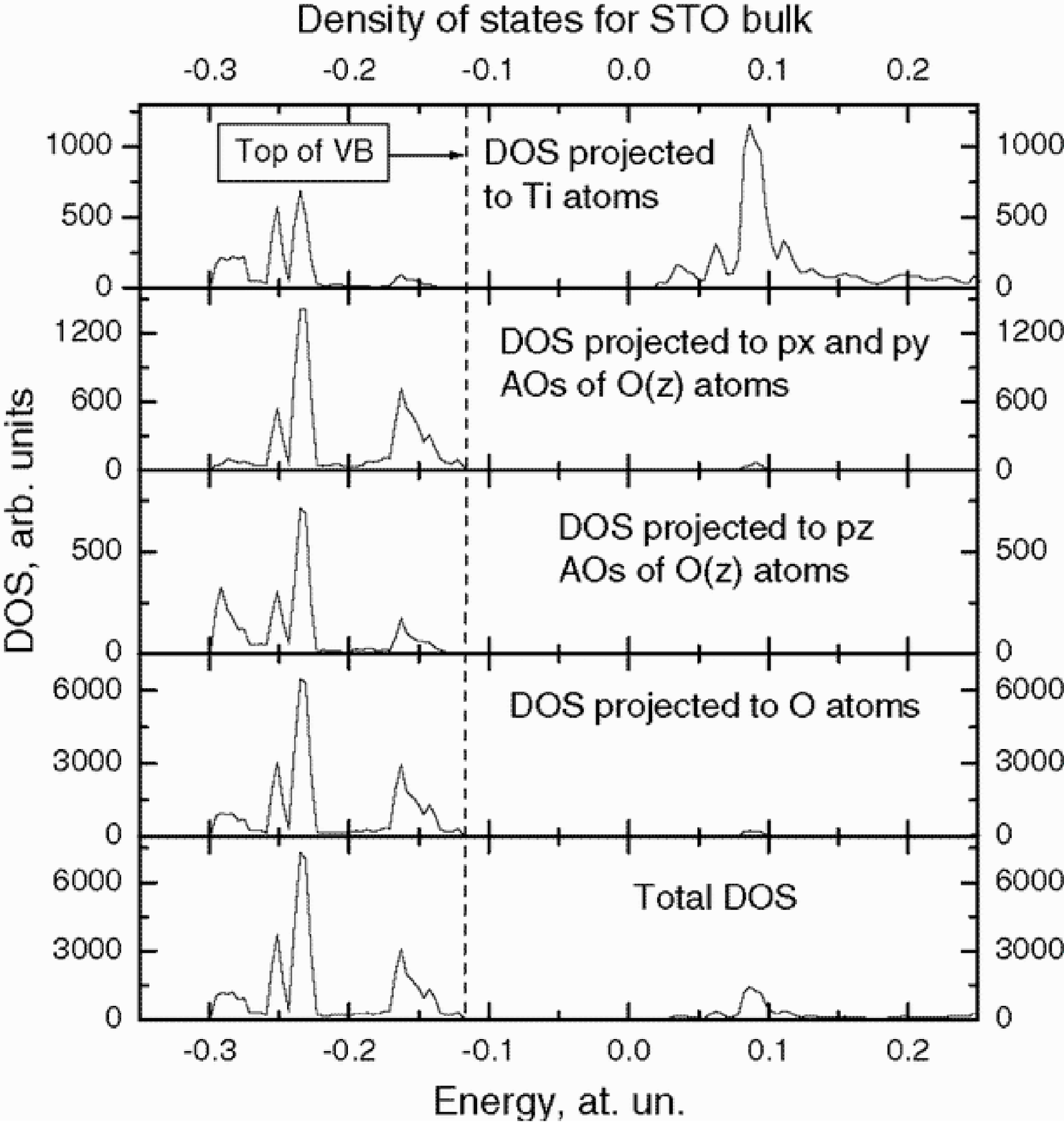,angle=0,width=15cm}
    \caption{
\small Total and projected DOS for the bulk STO.
      }
    \label{sto-1x1x4-dos}
  \end{center}
\end{figure}
\setlength{\fboxsep}{1pt}
\setlength{\fboxsep}{1pt}
\begin{figure}[htbp]
\def\baselinestretch{1.0}
\setlinespacing{1.0}
  \begin{center}
    \epsfig{file=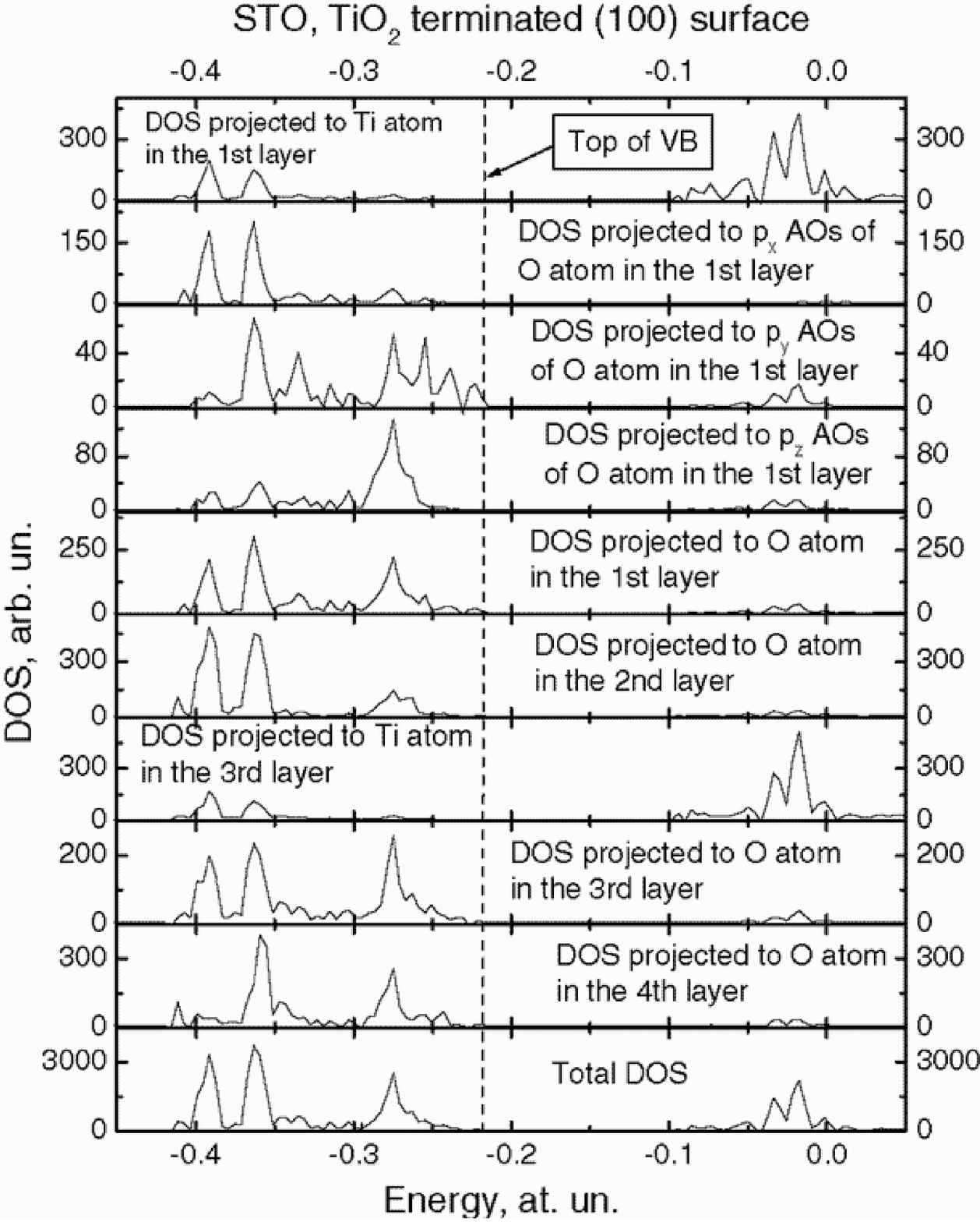,angle=0,width=15cm}
    \caption{
\small Total and projected DOS for the STO TiO$_{2}$-terminated
surface.
      }
    \label{sto-tio2-dos}
  \end{center}
\end{figure}
\setlength{\fboxsep}{1pt}
\begin{figure}[htbp]
\def\baselinestretch{1.0}
\setlinespacing{1.0}
  \begin{center}
    \epsfig{file=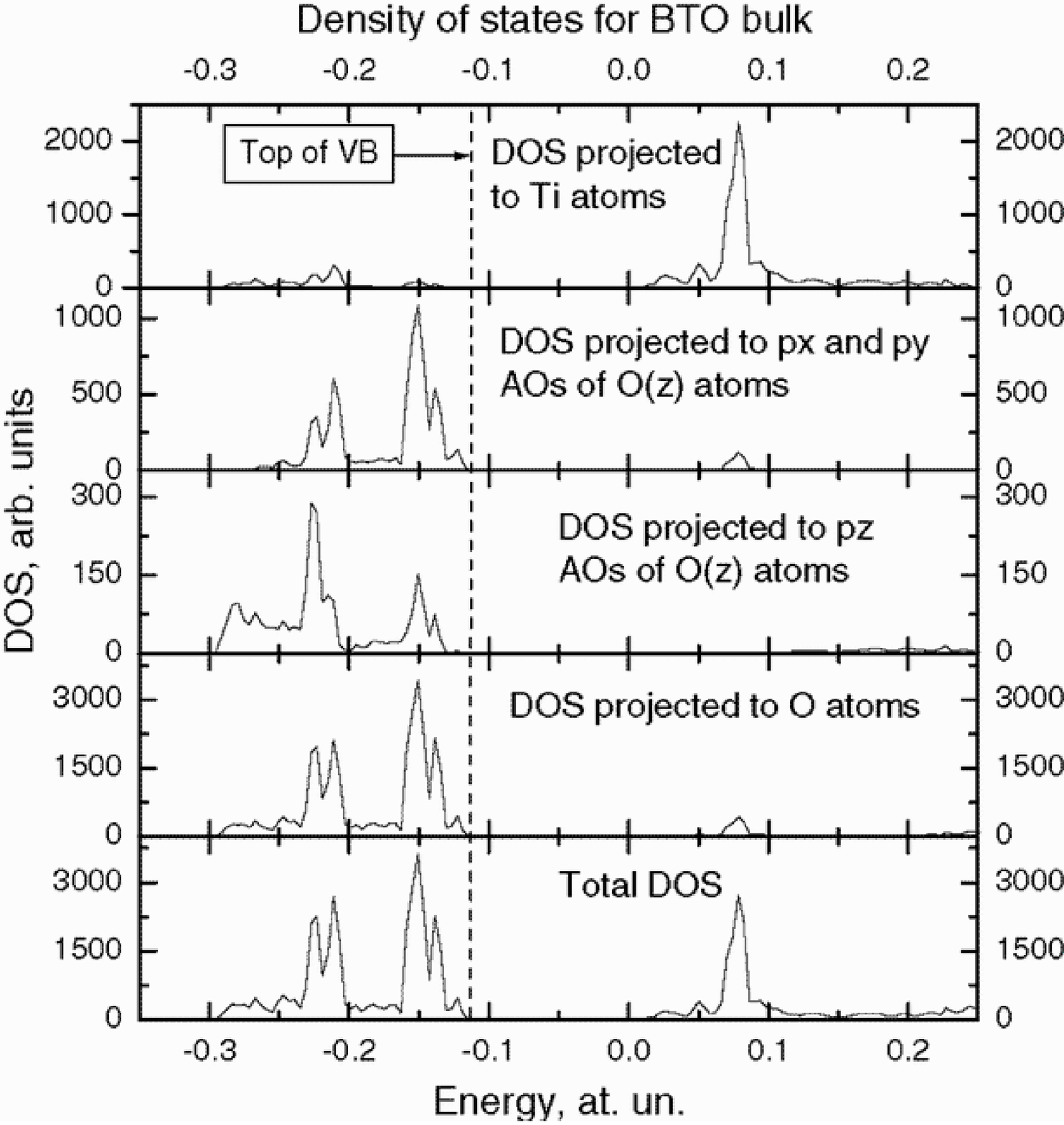,angle=0,width=15cm}
    \caption{
\small Total and projected DOS for the bulk BTO.
      }
    \label{bto-1x1x4-dos}
  \end{center}
\end{figure}
\setlength{\fboxsep}{1pt}
\begin{figure}[htbp]
\def\baselinestretch{1.0}
\setlinespacing{1.0}
  \begin{center}
    \epsfig{file=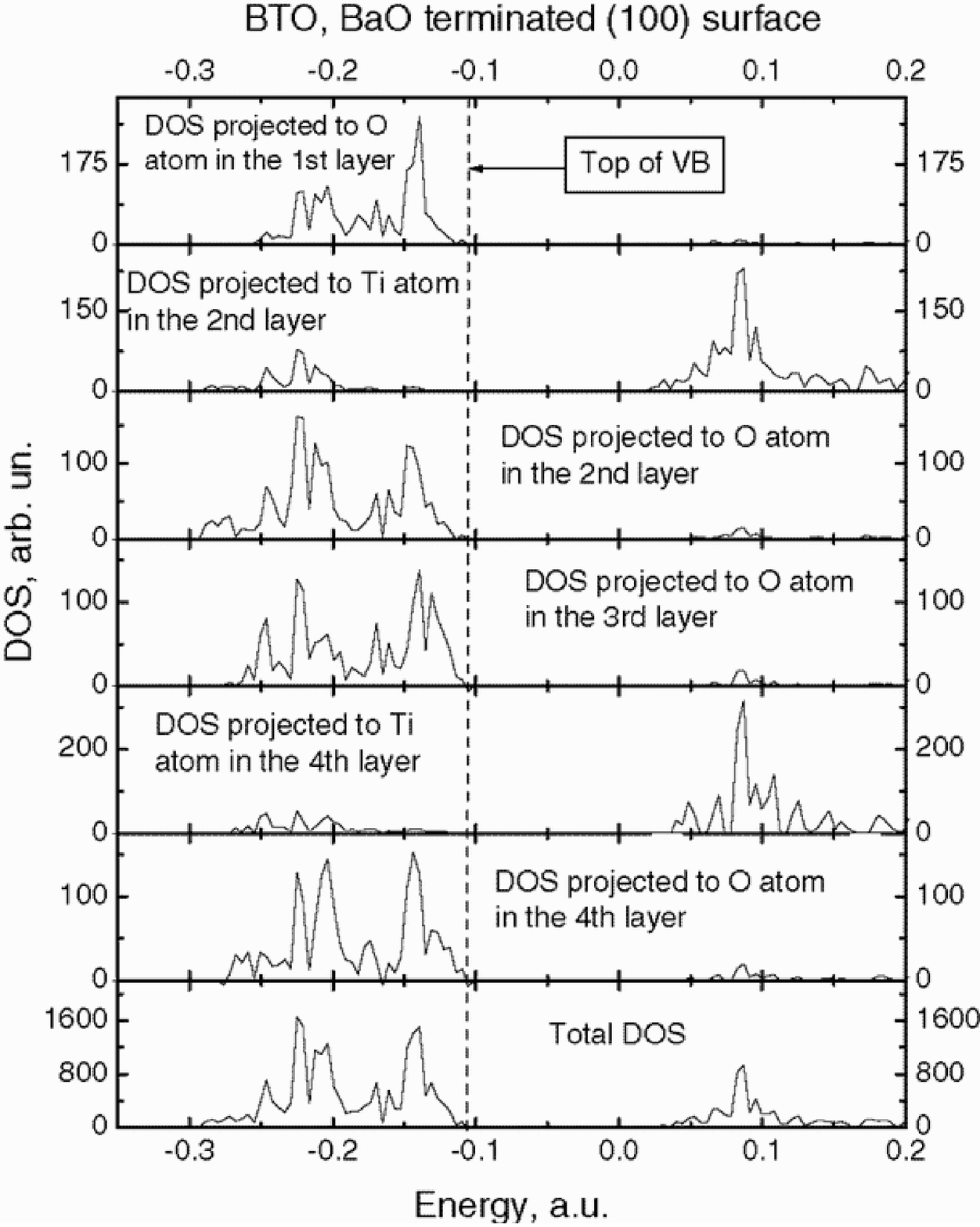,angle=0,width=15cm}
    \caption{
\small Total and projected DOS for the BaO-terminated surface.
      }
    \label{bto-bao-dos}
  \end{center}
\end{figure}
\setlength{\fboxsep}{1pt}
\begin{figure}[htbp]
\def\baselinestretch{1.0}
\setlinespacing{1.0}
  \begin{center}
    \epsfig{file=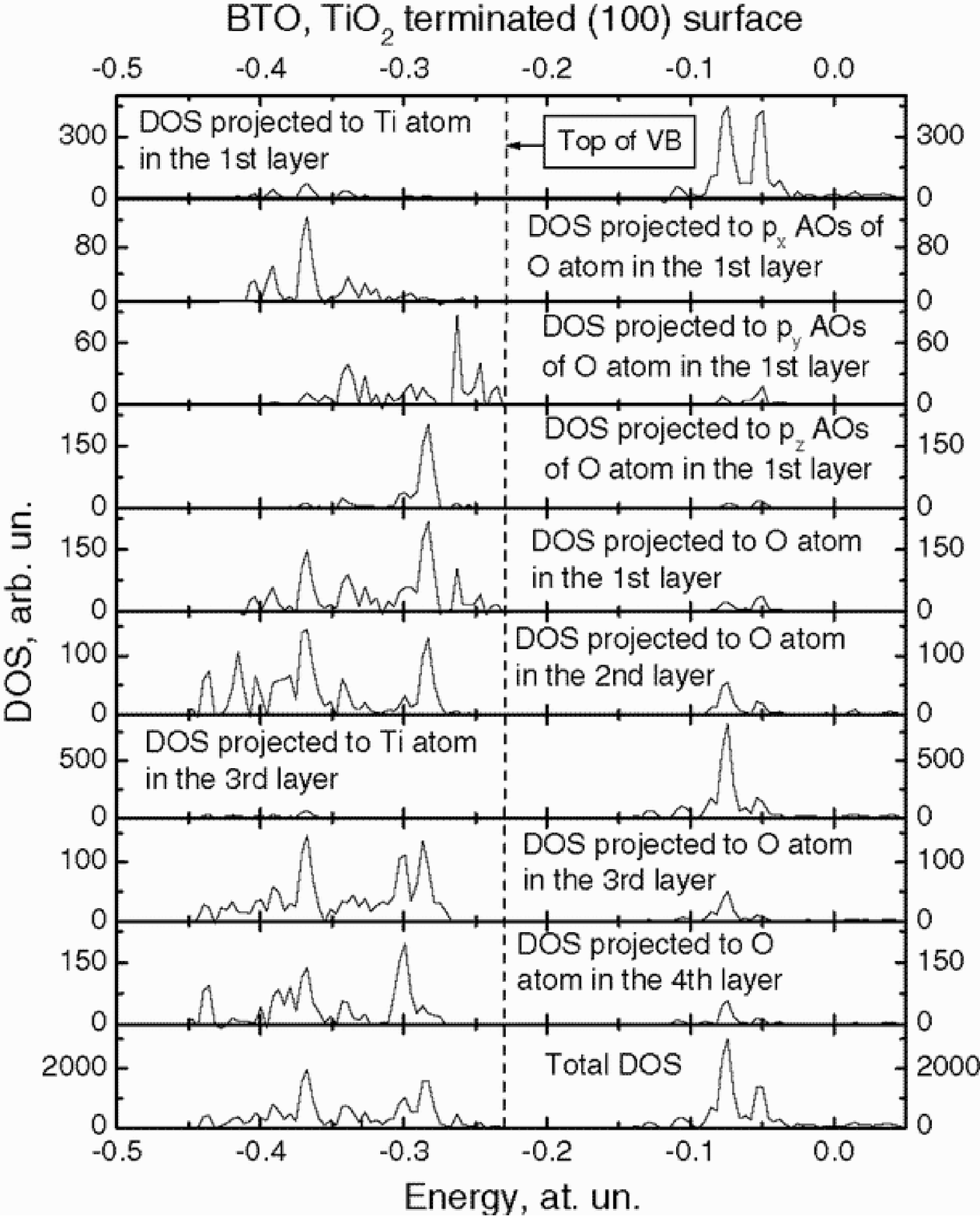,angle=0,width=15cm}
    \caption{
\small Total and projected DOS for the BTO TiO$_{2}$-terminated
surface.
      }
    \label{bto-tio2-dos}
  \end{center}
\end{figure}
\setlength{\fboxsep}{1pt}
\begin{figure}[htbp]
\def\baselinestretch{1.0}
\setlinespacing{1.0}
  \begin{center}
    \epsfig{file=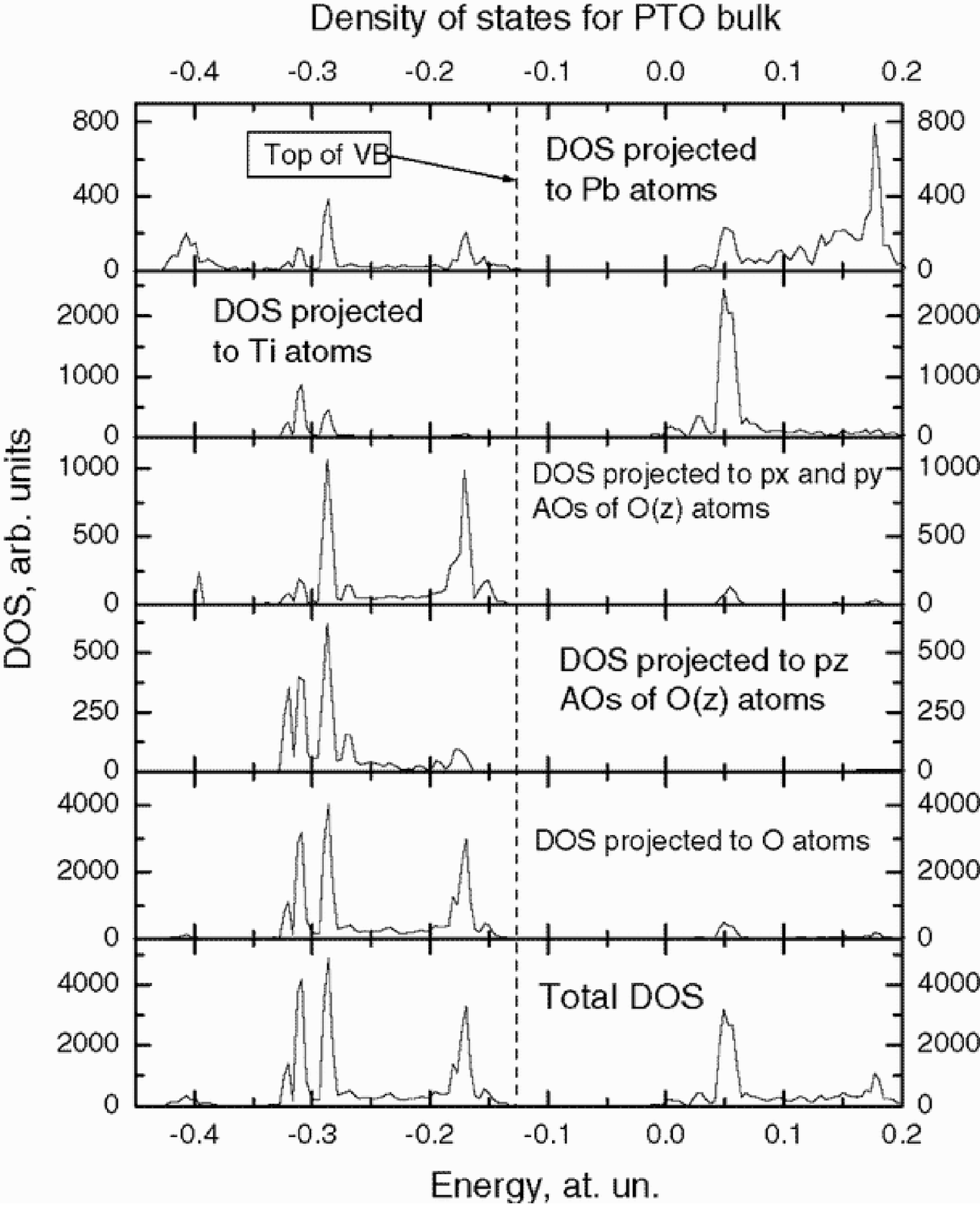,angle=0,width=15cm}
    \caption{
\small Total and projected DOS for the bulk PTO.
      }
    \label{pto-1x1x4-dos}
  \end{center}
\end{figure}
\setlength{\fboxsep}{1pt}
\begin{figure}[htbp]
\def\baselinestretch{1.0}
\setlinespacing{1.0}
  \begin{center}
    \epsfig{file=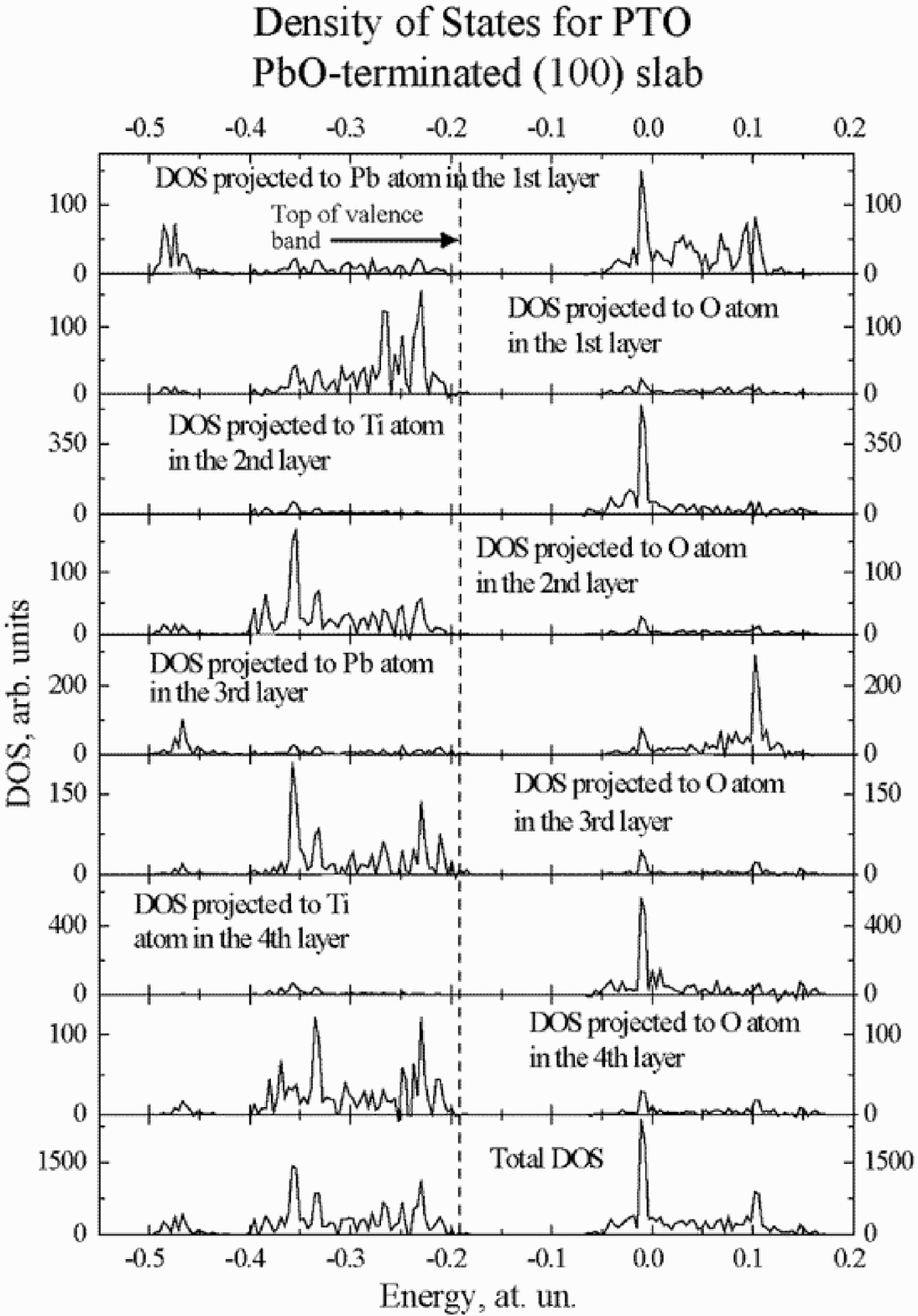,angle=0,width=15cm}
    \caption{
\small Total and projected DOS for the PbO-terminated surface.
      }
    \label{pto-pbo-dos}
  \end{center}
\end{figure}
\setlength{\fboxsep}{1pt}
\begin{figure}[htbp]
\def\baselinestretch{1.0}
\setlinespacing{1.0}
  \begin{center}
    \epsfig{file=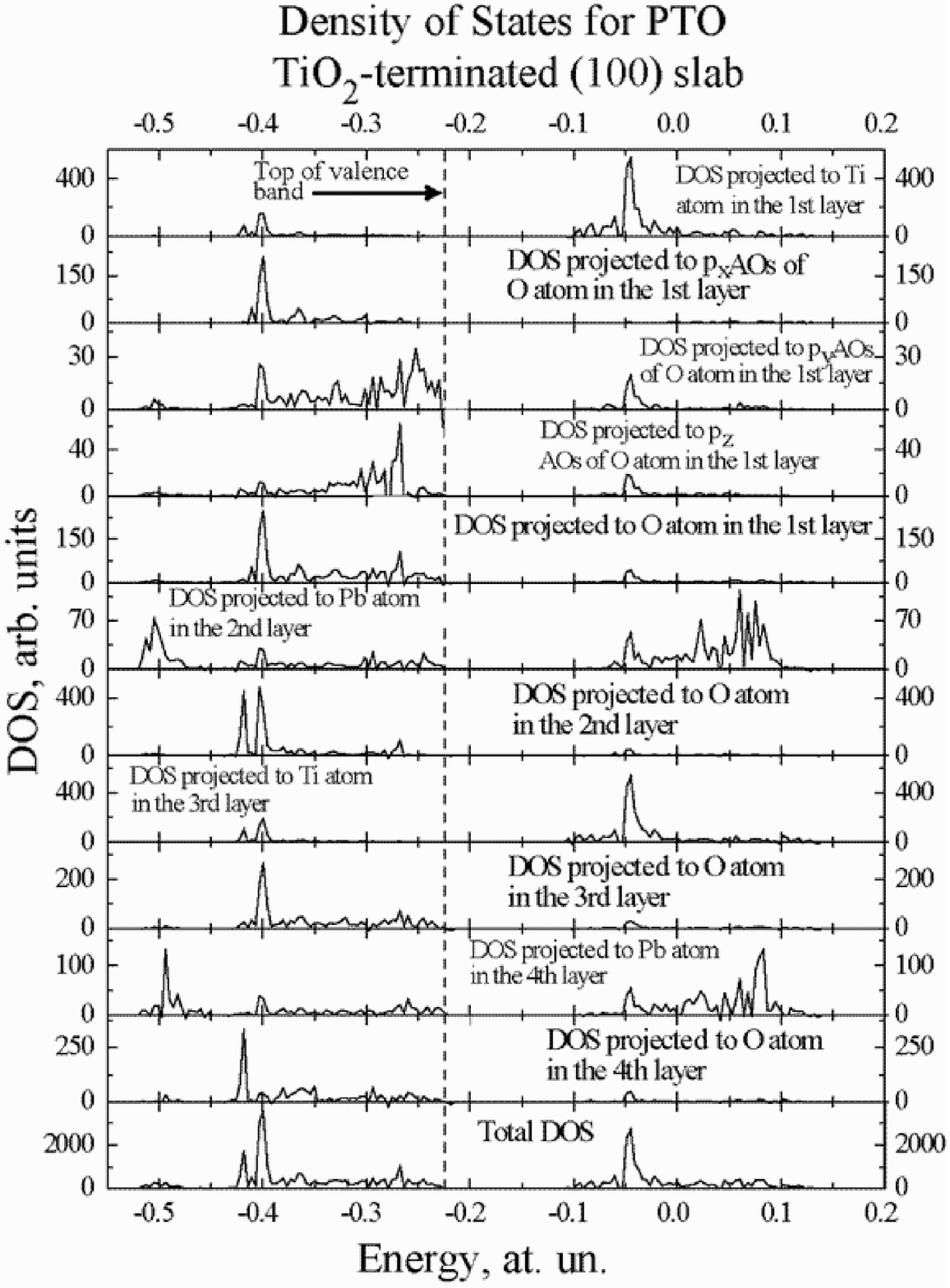,angle=0,width=15cm}
    \caption{
\small Total and projected DOS for the PTO TiO$_{2}$-terminated
surface.
      }
    \label{pto-tio2-dos}
  \end{center}
\end{figure}
The band structure calculated for another, TiO$_2$-terminated
surface of STO demonstrates less flatness of the VB top, in a
comparison with the SrO-termination. The indirect optical band gap
(M-$\Gamma$) 3.17 eV becomes by $\approx 0.46$ eV narrower as
compared with the bulk, \emph{i.e.} twice more reduced as the
SrO-terminated surface. For the TiO$_2$-terminated STO surface the
main contribution into the top of VB is made by O $2p_x$ and
$2p_y$ AOs which are perpendicular to Ti-O-Ti bridge (see Fig.
\ref{sto-tio2-dos}). The main contribution to the CB bottom comes
from the $3d$ AOs of Ti in third layer, the energy levels of
surface Ti lie a little bit higher in the energy. The calculated
STO DOS are in a good agreement with MIES and UPS spectra recently
measured for the TiO$_2$-terminated STO(001) surface
\cite{kempter-SS515}. Moreover, our calculated position of the VB
top for TiO$_2$-terminated STO with respect to the vacuum (5.9 eV)
practically coincides with the experimentally observed value of
$5.7\pm 0.2$ eV (Ref. \cite{kempter-SS515}).

The band structure calculated for the asymmetrically terminated
slab demonstrates the mixture of band structures obtained for the
two symmetrically terminated slabs discussed above. The STO band
gap becomes narrower (2.31 eV). The VB top consists mainly of O
$2p$ from TiO$_2$-terminated slab surface, and the CB bottom from
$3d$ AOs of Ti from a subsurface layer (II in Fig.
\ref{surfaces}(c)). The split of the upper VB ($\approx$ 0.8 eV)
is well pronounced in asymmetrical STO slab. The band structure of
the BTO(001) surfaces demonstrates practically the same behavior
as STO does (Fig. \ref{bto-band}, \ref{bto-1x1x4-dos},
\ref{bto-bao-dos}, \ref{bto-tio2-dos}, and Table \ref{gap}).
Nevertheless, the split of the VB upper band is pronounced more
for the TiO$_2$-terminated BTO, as compared with the STO surface.
Due to hybridization of Pb $6s$ and O $2p$ orbitals in PTO, the
calculated band structure and DOS slightly differ from those
calculated for STO and BTO (Fig. \ref{pto-band},
\ref{pto-1x1x4-dos}, \ref{pto-pbo-dos}, \ref{pto-tio2-dos}, and
Table \ref{gap}). The narrowest gap of both, bulk and surface band
structures corresponds to the transition between $\Gamma$ and X
points of the Brillouin zone. In the bulk, the VB top is formed
significantly by PB $6s$ AOs (which also make the main
contribution to the VB \emph{bottom}). The bulk CB bottom consist
of Ti $3d$ AOs, as in two other perovskites. Surprisingly, the
optical band gaps in PbO-terminated surfaces are not smaller, as
in BTO and STO, but even a little bit increases (up to 2.96 eV
with respect to 2.87 in the bulk). The VB top in the
PbO-terminated surface consists of a mixture of Pb $6s$ and O $2p$
AOs from a third layer, whereas the CB bottom is formed by Ti $3d$
AOs from a subsurface layer.

The VB top for the TiO$_2$-terminated PTO(001) surface at the X
point consists  of a mixture of the O $2p$ and Pb $6s$
AOs from \emph{both} surface and central layers. Moreover, the
main contribution comes from the orbitals of \emph{central} atoms.
The CB bottom for the TiO$_2$-terminated PTO consists mainly of
the Ti $3d$ AOs from a third layer.

The band structure calculated for the asymmetrical PTO slab shows
a mixture of the band structures of the two symmetrical slabs, as
it was observed for STO and BTO.

\section{Conclusions}

Using hybrid DFT-HF B3PW method, we calculated the surface
relaxation and the electronic structure of the two possible
terminations of the (001) surfaces for STO, BTO and PTO perovskite
crystals. The data obtained for the surface structure are in good
agreement with available results of theoretical \emph{ab initio}
calculations and with experimental data. The computed relaxed
surface energies are quite low, ~1-1.5 eV per unit cell. This
indicates that the surfaces with both termination are quite
stable, in agreement with ideas of a ``weak polarity"
\cite{noguerab,tasker-JPC-SSP12,noguera-SS365} and existing
experiments. The analysis of the atomic dipole moments shows that
the cations on the AO-terminated surfaces reveal the strong
electronic polarization along the $z$-axis. The calculated
difference electron density maps and bond populations demonstrate
an increase of the Ti-O bond covalency near the surfaces, and
additionally a weak covalency (polarization) of the Pb-O bond on
the PbO terminated surface.

The observed absence of the surface electronic states in the upper
VB for the AO-terminated (001) surfaces and considerable (~0.5 eV)
reduction of the gap for the TiO$_2$-terminated surface of all
three perovskites are important factor for the future treatment of
the electronic structure of surface defects on perovskite
surfaces, as well as adsorption and surface diffusion of atoms and
small molecules, which is relevant for catalysis, fuel cells, and
microelectronics.

\section*{Acknowledgements}

SP was partly supported by DFG. Authors are indebted to R.
Evarestov, G. Borstel, and R. Eglitis for numerous discussions.


\begin{thebibliography}{10}
\expandafter\ifx\csname url\endcsname\relax
  \def\url#1{\texttt{#1}}\fi
\expandafter\ifx\csname
urlprefix\endcsname\relax\def\urlprefix{URL }\fi

\bibitem{linesb}
M.~E. Lines, A.~M. Glass, Principles and Applications of
Ferroelectrics and
  Related Materials, Clarendon Press, Oxford, 1977.

\bibitem{noguerab}
C.~Noguera, Physics and Chemistry at Oxide Surfaces, Cambridge
University
  Press, New-York, 1996.

\bibitem{henricb}
V.~E. Henrick, P.~A. Cox, The Surface Science of Metal Oxides,
Cambridge
  University Press, New-York, 1994.

\bibitem{scott-feram}
J.~F. Scott, Ferroelectric Memories, Advanced Microelectronics 3,
Springer,
  Berlin, 2000.

\bibitem{bickel-PRL62}
N.~Bickel, G.~Schmidt, K.~Heinz, K.~M{\"{u}}ller, Ferroelectric
relaxation of
  the {S}r{T}i{O}$_3$(100) surface, Phys. Rev. Lett. 62~(17) (1989) 2009--2011.

\bibitem{hikita-ss287}
T.~Hikita, T.~Hanada, M.~Kudo, M.~Kawai, Structure and electronic
state of the
  {T}i{O}$_2$ and {S}r{O} terminated {S}r{T}i{O}$_3$(100) surfaces, Surf. Sci.
  287/288 (1993) 377--381.

\bibitem{ikeda-SS433}
A.~Ikeda, T.~Nishimura, T.~Morishita, Y.~Kido, Surface relaxation
and rumpling
  of {T}i{O}$_2$-terminated {S}r{T}i{O}$_3$(001) determined by medium energy
  ion scattering, Surf. Sci. 433-435 (1999) 520--524.

\bibitem{charlton-SS457}
G.~Charlton, S.~Brennan, C.~A. Muryn, R.~McGrath, D.~Norman, T.~S.
Turner,
  G.~Thorthon, Surface relaxation of {S}r{T}i{O}$_3$(001), Surf. Sci. 457
  (2000) L376--L380.

\bibitem{heide-SS473}
P.~A.~W. van~der Heide, Q.~D. Jiang, Y.~S. Kim, J.~W. Rabalais,
X-ray
  photoelectron spectroscopic and ion scattering study of the
  {S}r{T}i{O}$_3$(001) surface, Surf. Sci. 473 (2001) 59--70.

\bibitem{kempter-SS515}
W.~Maus-Friedrichs, M.~Frerichs, A.~Gunhold, S.~Krischok,
V.~Kempter,
  G.~Bihlmayer, The characterization of {S}r{T}i{O}$_3$(001) with {M}{I}{E}{S},
  {U}{P}{S}({H}e{I}) and first-principles calculations, Surf. Sci. 515 (2002)
  499--506.

\bibitem{kimura-prb51}
S.~Kimura, J.~Yamauchi, M.~Tsukada, S.~Watanabe, First-principles
study on
  electronic structure of the (001) surface of {S}r{T}i{O}$_3$, Phys. Rev. B
  51~(16) (1995) 11049--11054.

\bibitem{cohen-jpcs57}
R.~E. Cohen, Periodic slab {LAPW} computations for ferroelectric
  {B}a{T}i{O}$_3$, J. Phys. Chem. Solids 57~(10) (1996) 1393--1396.

\bibitem{cohen-frr194}
R.~E. Cohen, Surface effects in ferroelectrics: periodic slab
computations for
  {B}a{T}i{O}$_3$, Ferroelectrics 194 (1997) 323--342.

\bibitem{vand-prb56}
J.~Padilla, D.~Vanderbilt, \emph{Ab initio} study of {BaTiO}$_3$
surfaces,
  Phys. Rev. B 56~(3) (1997) 1625--1631.

\bibitem{vand-ss418}
J.~Padilla, D.~Vanderbilt, \emph{Ab initio} study of {SrTiO}$_3$
surfaces,
  Surf. Sci. 418 (1998) 64--70.

\bibitem{Vand-far114}
B.~Meyer, J.~Padilla, D.~Vanderbilt, Faraday Discussions 114: The
Surface
  Science of Metal Oxides, Royal Society of Chemistry, London, 1999, Ch. Theory
  of PbTiO$_3$, BaTiO$_3$, and SrTiO$_3$ surfaces, pp. 395--405.

\bibitem{Xue-ss550}
X.~Y. Xue, C.~L. Wang, W.~L. Zhong, The atomic and electronic
structure of the
  {TiO$_2$} and {BaO}-terminated {BaTiO$_3$}(001)surfaces in a paraelectric
  phase, Surf. Sci. 550 (2004) 73--78.

\bibitem{cora-catFD}
F.~Cora, C.~R.~A. Catlow, Faraday Discussions 114: The Surface
Science of Metal
  Oxides, Royal Society of Chemistry, London, 1999, Ch. {QM} investigations on
  perovskite-structured transition metal oxides: bulk, surfaces and interfaces,
  pp. 421--442.

\bibitem{Cheng-prb62}
C.~Cheng, K.~Kunc, M.~H. Lee, Structural relaxation and
longitudinal dipole
  moment of {SrTiO}$_3$(001) (1$\times$1) surface, Phys. Rev. B 62~(15) (2000)
  10409--10418.

\bibitem{Tinte-aip535}
S.~Tinte, M.~D. Stachiotti, Atomistic simulation of surface
effects in
  {BaTiO}$_3$, AIP. conf. proc. 535 (2000) 273--282.

\bibitem{Tinte-prb64}
S.~Tinte, M.~D. Stachiotti, Surface effects and ferroelectric
phase transitions
  in {BaTiO}$_3$ ultrathin films, Phys. Rev. B 64 (2001) 235403.

\bibitem{Pisani-ed-96}
C.~Pisani (Ed.), Quantum-Mechanical \emph{Ab-initio} Calculations
of the
  Properties of Crystalline Materials, Vol.~67 of Lecture Notes in Chemistry,
  Springer, Berlin, 1996.

\bibitem{curtiss-JCP106}
L.~A. Curtiss, K.~Raghavachari, P.~C. Redfern, J.~A. Pople,
Assessment of
  {G}aussian-2 and density functional theories for the computation of entalpies
  of formation, J. Chem. Phys. 106~(3) (1997) 1063--1079.

\bibitem{curtiss-JCP109}
L.~A. Curtiss, P.~C. Redfern, K.~Raghavachari, J.~A. Pople,
Assessment of
  {G}aussian-2 and density functional theories for the computation of
  ionization potentials and electron affinities, J. Chem. Phys. 109~(1) (1998)
  42--55.

\bibitem{harrison-b3lyp}
J.~Muscat, A.~Wander, N.~M. Harrison, On the prediction of band
gaps from
  hybrid functional theory, Chem. Phys. Lett. 342 (2001) 397--401.

\bibitem{pisk-BS}
S.~Piskunov, E.~Heifets, R.~I. Eglitis, G.~Borstel, Bulk
properties and
  electronic structure of {SrTiO}$_3$, {BaTiO}$_3$ and {PbTiO}$_3$ perovskites:
  an \emph{ab initio} {HF/DFT} study, Comp. Mat. Sci. 29 (2004) 165--178.

\bibitem{heif-ss02}
E.~Heifets, R.~I. Eglitis, E.~A. Kotomin, J.~Maier, G.~Borstel,
  First-principles calculations for {SrTiO}$_3$(100) surface structure, Surf.
  Sci. 513 (2002) 211--220.

\bibitem{heifss2000}
E.~Heifets, E.~A. Kotomin, J.~Maier, Semi-empirical simulations of
surface
  relaxation for perovskite titanates, Surf. Sci. 462 (2000) 19--35.

\bibitem{heif-prb01}
E.~Heifets, R.~I. Eglitis, E.~A. Kotomin, J.~Maier, G.~Borstel,
\emph{Ab
  initio} modeling of surface structure for {SrTiO}$_3$ perovskite crystals,
  Phys. Rev. B 64 (2001) 235417.

\bibitem{kotomin-TSF400}
E.~A. Kotomin, R.~I. Eglitis, J.~Maier, E.~Heifets, Calculations
of the atomic
  and electronic structure for {SrTiO}$_3$ perovskite thin films, Thin Solid
  Films 400 (2001) 76--80.

\bibitem{borstel-PSSb236}
G.~Borstel, R.~I. Eglitis, E.~A. Kotomin, E.~Heifets, Modelling of
defects and
  surfaces in perovskite ferroelectrics, Phys. Stat. Sol. (b) 236~(2) (2003)
  253--264.

\bibitem{CRman}
V.~R. Saunders, R.~Dovesi, C.~Roetti, M.~Causa, N.~M. Harrison,
R.~Orlando,
  C.~M. Zicovich-Wilson, CRYSTAL'98 User's Manual, Universita di Torino, Torino
  (1998).

\bibitem{CR-http1}
http://www.chimifm.unito.it/teorica/crystal/crystal.html.

\bibitem{CR-http2}
http://www.cse.clrc.ac.uk/cmg/crystal.

\bibitem{becke-hybr}
A.~D. Becke, Density-functional thermochemistry. {III}. the role
of exact
  exchange, J. Chem. Phys. 98~(7) (1993) 5648--5652.

\bibitem{PW1}
J.~P. Perdew, Y.~Wang, Accurate and simple density functional for
the
  electronic exchange energy: {G}eneralized gradient approximation, Phys. Rev.
  B 33~(12) (1986) 8800--8802.

\bibitem{PW2}
J.~P. Perdew, Y.~Wang, Erratum: Accurate and simple density
functional for the
  electronic exchange energy: {G}eneralized gradient approximation, Phys. Rev.
  B 40~(5) (1989) 3399.

\bibitem{PW3}
J.~P. Perdew, Y.~Wang, Accurate and simple analytic representation
of the
  electron-gas correlation energy, Phys. Rev. B 45~(23) (1992) 13244--13249.

\bibitem{hw1}
P.~J. Hay, W.~R. Wadt, \emph{Ab initio} effective core potentials
for molecular
  calculations. {P}otentials for the transition metal atoms {S}c to {H}g, J.
  Chem. Phys. 82~(1) (1984) 270--283.

\bibitem{hw2}
P.~J. Hay, W.~R. Wadt, \emph{Ab initio} effective core potentials
for molecular
  calculations. {P}otentials for main group elements {N}a to {B}i, J. Chem.
  Phys. 82~(1) (1984) 284--298.

\bibitem{hw3}
P.~J. Hay, W.~R. Wadt, \emph{Ab initio} effective core potentials
for molecular
  calculations. {P}otentials for {K} to {A}u including the outermost core
  orbitals, J. Chem. Phys. 82~(1) (1984) 299--310.

\bibitem{monkhorst}
H.~J. Monkhorst, J.~D. Pack, Special points for {B}rillouin-zone
integrations,
  Phys. Rev. B 13~(12) (1976) 5188--5192.

\bibitem{bredov-evarestov}
T.~Bredow, R.~A. Evarestov, K.~Jug, Implementation of the {C}yclic
{C}luster
  {M}odel in {H}artree-{F}ock {LCAO} calculations of crystalline systems, Phys.
  Stat. Solidi. (b) 222 (2000) 495--516.

\bibitem{springIII}
K.~H. Hellwege, A.~M. Hellwege (Eds.), Ferroelectrics and Related
Substances,
  Vol.~3 of New Series, Landolt-Bornstein, Springer Verlag, Berlin, 1969, group
  III.

\bibitem{shirane}
B.~G. Shirane, R.~Repinsky, B.~C. Frazer, X-ray and neutron
diffraction study
  of ferroelectric {PbTiO}$_3$, Acta. Cryst. 9 (1956) 131--140.

\bibitem{num-rec-f77}
W.~H. Press, S.~A. Teukolsky, W.~T. Vetterling, B.~P. Flannery,
Numerical
  Recipies in Fortran77, 2nd Edition, Cambridge Univ. Press, Cambridge, MA,
  1997.

\bibitem{tasker-JPC-SSP12}
P.~W. Tasker, The stability of ionic crystal surfaces, J. Phys. C:
Solid State
  Phys. 12 (1979) 4977--4984.

\bibitem{noguera-SS365}
J.~Goniakowski, C.~Noguera, The concept of weak polarity: an
application to the
  {SrTiO}$_3$(001) surface, Surf. Sci. 365 (1996) L657--L662.

\bibitem{evar-WTAO}
R.~A. Evarestov, V.~P. Smirnov, D.~E. Usvyat, Local properties of
the
  electronic structure of cubic {SrTiO}$_3$, {BaTiO}$_3$ and {PbTiO}$_3$
  crystals in {W}annier-type function approach, is submitted to Solid State
  Communications (2003).

\bibitem{wemple}
S.~H. Wemple, Polarization fluctuations and the optical-absorption
edge in
  {BaTiO}$_3$, Phys. Rev. B 2~(7) (1970) 2679--2689.

\bibitem{peng-chang}
C.~H. Peng, J.~F. Chang, S.~Desu, Mater. Res. Soc. Symp. Proc. 243
(1992) 21.

\bibitem{benthem}
K.~van Benthem, C.~Elsaesser, R.~H. French, Bulk electronic
structure of
  {SrTiO}$_3$: {E}xperiment and theory, J. Appl. Phys. 90~(12) (2001)
  6156--6164.

\bibitem{lasaro}
S.~de Lasaro, E.~Longo, J.~R. Sambrano, A.~Beltran, Structural and
electronic properties of {PbTiO}$_3$ slabs: a DFT periodic study,
Surf. Sci.  552 (2004) 149--159.

\end{thebibliography}
\end{document}